\documentclass[preprint,12pt]{elsarticle}

\usepackage{amssymb}
\usepackage{xcolor}
\usepackage{caption}
\usepackage{textcomp} 

\usepackage{lineno}

\journal{Biofilm}

\begin{document}
\begin{frontmatter}

\title{Latent Space-Driven Quantification of Biofilm Formation using Time-Resolved Droplet Microfluidics}

\author[inst1,$^1$]{Daniela Pérez Guerrero}
\author[inst1,inst2,$^1$]{Jesús Manuel Antúnez Domínguez}
\author[inst2]{Aurélie Vigne}
\author[inst1]{Daniel Midtvedt}
\author[inst3,inst4,inst5]{Wylie Ahmed}
\author[inst2]{Lisa D. Muiznieks}
\author[inst1,inst6]{Giovanni Volpe}
\author[inst1]{Caroline Beck Adiels\corref{cor1}}

\fntext[$^1$]{These authors contributed equally to this work.}
\cortext[cor1]{Corresponding author.}
\ead{caroline.adiels@physics.gu.se}

\affiliation[inst1]{organization={University of Gothenburg, Physics Department,},
            addressline={ Origovägen 6 B}, 
            city={Gothenburg},
            postcode={412 58}, 
            country={Sweden}}

\affiliation[inst2]{organization={Microfluidics Innovation Center},
            addressline={172 Rue de Charonne}, 
            city={Paris},
            postcode={75011}, 
            country={France}}

\affiliation[inst3]{organization={Laboratoire de Physique Théorique (LPT), Université de Toulouse, CNRS, UPS},
            addressline={118 Rte de Narbonne}, 
            city={Toulouse},
            postcode={31062}, 
            country={France}}

\affiliation[inst4]{organization={Molecular, Cellular and Developmental biology unit (MCD), Centre de Biologie Intégrative (CBI), Université de Toulouse, CNRS, UPS},
            addressline={118 Rte de Narbonne}, 
            city={Toulouse},
            postcode={31062}, 
            country={France}}
            
\affiliation[inst5]{organization={Department of Physics, California State University Fullerton},
            addressline={800 N. State College Blvd}, 
            postcode={92831}, 
            state={CA},
            country={USA}}
            
\affiliation[inst6]{organization={Science for Life Laboratory, Physics Department, University of Gothenburg},
            city={Gothenburg},
            country={Sweden}}

\begin{abstract}
Bacterial biofilms play a significant role in various fields that impact our daily lives, from detrimental public health hazards to beneficial applications in bioremediation, biodegradation, and wastewater treatment. However, high-resolution tools for studying their dynamic responses to environmental changes and collective cellular behavior remain scarce. To characterize and quantify biofilm development, we present a droplet-based microfluidic platform combined with an image analysis tool for \textit{\textit{in-situ}} studies. In this setup, \textit{Bacillus subtilis} was inoculated in liquid Lysogeny Broth microdroplets, and biofilm formation was examined within emulsions at the water-oil interface. Bacteria were encapsulated in droplets, which were then trapped in compartments, allowing continuous optical access throughout biofilm formation. Droplets, each forming a distinct microenvironment, were generated at high throughput using flow-controlled pressure pumps, ensuring monodispersity. A microfluidic multi-injection valve enabled rapid switching of encapsulation conditions without disrupting droplet generation, allowing side-by-side comparison. Our platform supports fluorescence microscopy imaging and quantitative analysis of droplet content, along with time-lapse bright-field microscopy for dynamic observations. To process high-throughput, complex data, we integrated an automated, unsupervised image analysis tool based on a Variational Autoencoder (VAE). This AI-driven approach efficiently captured biofilm structures in a latent space, enabling detailed pattern recognition and analysis. Our results demonstrate the accurate detection and quantification of biofilms using thresholding and masking applied to latent space representations, enabling the precise measurement of biofilm and aggregate areas. By integrating AI-driven analysis with droplet microfluidics, our platform offers a scalable and robust tool for advancing strategies in both biofilm applications and control.
\end{abstract}

\begin{graphicalabstract}
\begin{center}
\includegraphics[scale = 0.73]{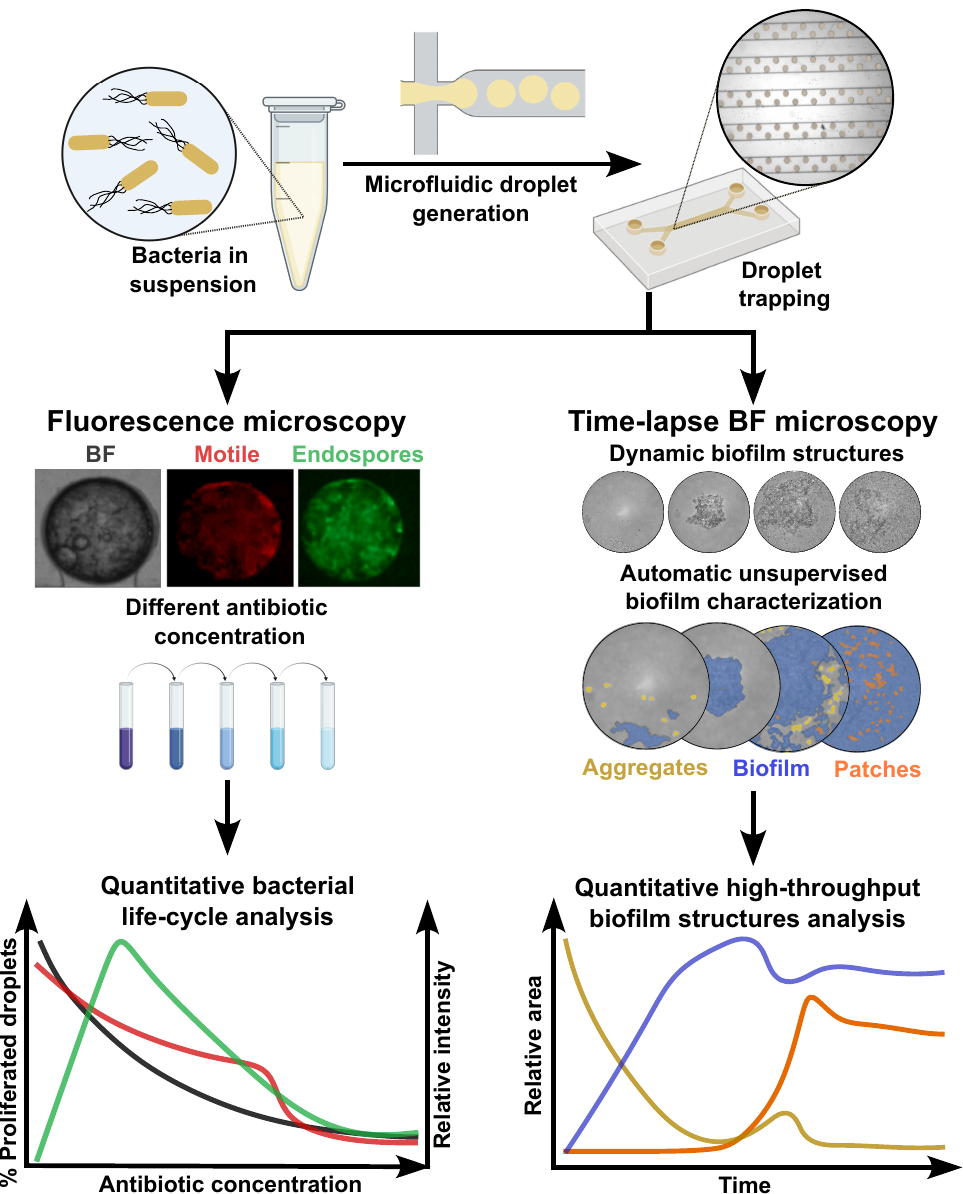}
\end{center}
\end{graphicalabstract}

\begin{highlights}
\item A droplet-based microfluidic platform generates bacteria-containing monodisperse droplets via pressure-controlled flow, enabling reproducible, \textit{in-situ} biofilm analysis.
\item This platform is used in conjunction with fluorescent and brightfield microscopy and is adapted to different experiments according to the data acquired, time resolution, and throughput.
\item A Variational Autoencoder uses an unsupervised approach to identify biofilm structures in the latent space, enabling precise and automated quantification of biofilm formation.
\item Time-resolved dynamics of biofilm behavior allow detailed descriptions of biofilm formation and growth according to environmental conditions.
\end{highlights}


\end{frontmatter}

\section{Introduction}
\label{sec:sample1}

Bacteria, despite their simplicity, are highly adaptable organisms that rely on complex collective behaviors to withstand hostile environmental conditions \cite{Popat2015}. These abilities emerge from their ability to move through their surroundings \cite{Mitchell2006,Wadhwa2022} and communicate via quorum sensing \cite{Miller2001}, allowing for association and synchronization.
Biofilm formation is one of the most prominent examples of such collective behavior. In this state, cells embed themselves in an extracellular matrix (ECM) that anchors them to the interface between two media \cite{Parsek2005}. The biofilm structure promotes close cell-to-cell interactions and task differentiation, such as the emergence of motile scouts or recalcitrant cells like endospores \cite{Arnaouteli2021}. This organization enhances bacterial resistance to environmental stressors, including predation \cite{Wucher2021}, desiccation \cite{Espinal2012}, and antibiotics \cite{Hathroubi2017}. In the case of antibiotics, biofilms contribute to resistance by facilitating genetic exchange and promoting  collective tolerance \cite{Perez2020, Sun2019}. While they present challenges in healthcare (e.g., persistent infections and sterilization difficulties \cite{Kolpen2022}), biofilms also have beneficial applications in bioremediation and biotechnology \cite{Sonawane2022} due to their unique metabolic capabilities.

Most experimental biofilm studies rely on bulk culture systems like agar and well plates, or on substrate-based systems analyzed \textit{ex-situ}, such as the Modified Robbins Device \cite{Linton1999} or the Drip flow biofilm reactor \cite{Schwartz2010}. However, these methods disrupt the original environment and prevent real-time observation \cite{Azeredo2017}. In contrast, microfluidics offers an alternative platform for microbiological studies \cite{Kaminski2016a} enabling \textit{in situ} visualization through transparent microfluidic chips. These systems also improve experimental efficiency by minimizing handling, reducing the use of costly or hazardous reagents, and allowing precise control through engineered microstructures \cite{Niculescu2021}. When designed with two immiscible liquid phases, microfluidic systems can produce droplets—discrete, compartmentalized environments that mimic isolated microhabitats. Each droplet functions as an individual experimental unit within an emulsion-like suspension, allowing for parallelization and high-throughput data generation \cite{Chen2022,Jiang2022}.

In our study, we utilize this capability by inoculating biofilm-forming bacteria in the aqueous phase of microfluidic droplets, enabling high-throughput imaging of self-contained microenvironments. Each droplet acts as a miniature culture, from which we can collect detailed microscopy data across thousands of droplets. To efficiently process and analyze the large and complex image dataset, we employ Variational Autoencoders (VAEs)\cite{kingma2014auto, Volpe2025}, which offer an unsupervised approach for learning and representing intricate image structures without requiring extensive labeled datasets. A VAE consists of an encoder, which compresses each input image crop into a low-dimensional latent representation (one- or multi-dimensional), and a decoder trained to reconstruct the original crop from this latent space. 

However, our approach deviates from the conventional use of VAEs: we bypass the decoder entirely. Rather than reconstructing images, we work directly in the latent space, which captures essential spatial and structural features. This enables automated segmentation and extraction of quantitative data directly from input images. While this is not the standard application of VAEs, it offers a novel, efficient and scalable strategy to reproducible image interpretation—particularly well-suited for high-throughput datasets.

We present a workflow that integrates droplet-based microfluidics with this unsupervised VAE-based image analysis to monitor and quantify the biofilm formation process of \textit{Bacillus subtilis}. The microfluidic platform is compatible with various microscopy modalities, including fluorescence and bright-field microscopy, with temporal resolution tailored to match the droplet generation throughput. The VAE efficiently processes the large volume of image data collected across droplets and time points, enabling the automated identification of biofilms and bacterial aggregates. 
By analyzing spatial patterns within the latent-space, our approach allows for quantitative tracking of biofilm dynamics over time and under different environmental conditions—such as nutrient availability and chemical treatments—revealing factors that promote or inhibit biofilm formation. Overall, our method provides a robust, scalable solution for high-throughput, quantitative studies of biofilm behavior by combining droplet microfluidics with unsupervised deep learning.

\section{Materials and methods}

\subsection{Bacterial Strain}

The genetically modified fluorescent strain \textit{Bacillus subtilis} MTC871 (modifications stated in SI Section~1 \textit{Bacillus subtilis} growth and calibration curves) was used as the model organism. This strain exhibits red fluorescence when motile and green fluorescence when forming endospores. The strain was grown in Lysogeny Broth Lennox medium (Sigma-Aldrich, Germany) at the optimal growth temperature \cite{Gu2019}, 37\textdegree C, until the $OD_{600}$ reading of the culture was stable ($OD_{600}$ = 1.0). To assess the reproducibility and repeatability of the experiments, the growth curve of \textit{Bacillus subtilis} and the relation between the OD of a sample and bacterial concentration were established with the spread plate technique (SI Fig.~1: \textit{Bacillus subtilis} growth curve and Fig.~2: \textit{Bacillus subtilis} calibration curve, respectively).

The bacteria were inoculated in Falcon reservoirs filled with Lysogeny Broth medium. One falcon reservoir was kept sterile as a negative control to generate empty droplets. \textit{Bacillus subtilis} (in a concentration of $OD_{600}$ = $0.06$) was added in the rest. Reagents were added to assess their impact on biofilm formation. The effects of Penicillin/Streptomycin (Pen/Strep, $5000$ units of penicillin and $5~mg$ streptomycin per  $1~mL$ , Sigma-Aldrich, Germany) at concentrations ranging from 0 to $4~\mu$L per $1~mL$ of medium, as well as filter-sterilized glycerol (Dutscher, France) at a fixed concentration of $1\%$ (v/v), were investigated within the droplets.
 
\subsection{Droplet-based microfluidic platform}

Each droplet functions as a self-contained microenvironment, making it essential that all components involved—surfactant, oil, aqueous phase, chip substrate, and analysis techniques—were carefully selected and tuned in harmony. In this context, where the species of interest were encapsulated within individual droplets, the biocompatibility of every element becomes crucial.
The continuous phase consisted of HFE-7500 perfluorinated oil (Novec7500, 3M) with a $1\%$ (w/w) concentration of Fluorosurf surfactant (Emulseo, France). The negative control, bacterial suspensions, and continuous phase reservoirs were connected to a pressure pump OB1 (Elveflow, France) to accurately control the liquid injection into the chip and coupled with flow sensors, MFS2D (Elveflow, France) for the dispersed phase and MFS3D (Elveflow, France) for the continuous phase. A diagram of the complete setup can be found in SI Section~2: Experimental setup. The pressure pump can establish a feedback loop to allow precise control of the liquid flow rate. Teflon tubing with an outer diameter of $1/32''$ (Darwin Microfluidics, France) was used to connect all the fluidic parts. To properly control the dispersed phase, $20~cm$ of $100~\mu m$ resistance tubing (Elveflow, France) was added to the platform after the flow sensor. A rotary multi-injection valve (MUX distributor from Elveflow, France) was added to allow the permutation of the dispersed solutions (negative control and bacterial suspensions) without disrupting the droplet production process. The sequence of measured solutions was spaced with pure perfluorinated oil included in the multi-injection valve channels to prevent the mixing of consecutive solutions, as shown in SI Fig.~3A: Sequence.

\subsection{Microfluidic chip}

Monodispersity of the droplets was the main control parameter to ensure that all the droplets were equivalent. Coupled with the VAE analysis tool, high-throughput droplet production provides the same volume of information as hundreds of bulk experiments from a single batch \cite{Liu2023}. The generated droplets were produced with a bacterial suspension of $OD_{600}$ = $0.06$ and had a volume of approximately $1.7$ $nL$. Thus, according to the obtained proportion, the CFU (Colony Forming Units) per droplet was $14$. This number was considered an approximation of the number of bacteria per droplet after encapsulation. 

A microfluidic chip (Fluidic 719, Microfluidic Chipshop, Germany) made of the cyclo-olefin copolymer Topas was used. The continuous and dispersed phases were perfused through different inlets, using a flow-focusing geometry with a nozzle of $82~\mu m$ to produce droplets as seen in SI Fig.~3B: Droplet generation. The droplets were released into a serpentine channel with storage positions. Prior to the experiment, the channels were treated with Aquapel \cite{Feng2022} and washed with the continuous phase to ensure the dispersed phase dewets, thereby preventing droplets from coalescing if they come into contact with each other. The generation of droplets was controlled through flow rate, with the continuous phase at $35~\mu L/min$ and the solutions at $2~\mu L/min$. A droplet size of $150~\mu m$ diameter was generated with high monodispersity (CV = 2\%), as shown in SI Fig.~4: Droplet size distribution.

Along the serpentine channel, which has a height of $110~\mu m$, a total of $2261$ droplet traps were alternatively positioned on each side. Each trap has a diameter of $173~\mu m$ and an additional height of $80~\mu$m. The aqueous phase droplets, less dense than the perfluorinated oil, floated in the serpentine channel and became stuck in the traps, preventing their displacement, as illustrated in SI Fig.~3C: Droplet trapping. Given the size of the droplets, only one droplet could be contained in each trap, and the rest dodged it and continued their trajectory until they reached the next free trap or the chip outlet. However, droplets could be replaced or pushed out from their trap at high flow rates. The dispersed phase flow was stopped once the reagent sequence was finished, while the continuous phase flow continued until all the droplets were trapped. Once the desired number of traps was filled, low-volume displacement plugs (Microfluidic Chipshop, Germany) were used to seal the inlets, preventing evaporation. 

\subsection{Biofilm observation}

The chip architecture provided stationary and stable bacteria-containing droplets, which were monitored over time as shown in Fig.~\ref{fig:fig_1}A. Intra-droplet pellicle formation was triggered in the oil-liquid interfaces of the droplet surface due to the high oxygen permeability of the oil. Further experimental evidence of this phenomenon can be found in SI Section~4, Perfluorinated oil oxygen permeability evaluation. This process was monitored using fluorescence or bright-field microscopy. For fluorescence microscopy measurements, $2000$ droplets were imaged using the specific filters Texas Red, targeting the fluorescence of motile cells (\textit{mKate} fluorophore linked to flagellin production) with peak excitation/emission wavelengths of $572/629$ $nm$, and FITC, targeting endospore formation (\textit{citrus} fluorophore linked to PsspB protein production) with peak excitation/emission wavelengths of $500/535$ $nm$.

Two different approaches were used for brightfield microscopy measurements. First, a low-throughput time-lapse of a few droplets with a short acquisition interval was used to establish an analysis model and reference for the dynamics of biofilm formation. The images were acquired using an inverted microscope (DMi 6000B; Leica Microsystems, Wetzlar, Germany) equipped with a SOLA III U-nIR LED source (Lumencor, Beaverton, USA) and a Hamamatsu Orca Flash $4$ camera (Hamamatsu Photonics, Kista, Stockholm). Six droplets were automatically imaged at $63x$ magnification every $7$ minutes for $17.5$ hours. Second, to assess the utility of the developed platform, high-throughput time-lapse measurements were conducted using numerous droplets with extended acquisition intervals.

This approach was applied to evaluate the platform’s ability to capture glycerol’s established role as a biofilm promoter. A total of $676$ trapped droplets were manually imaged at defined time intervals to monitor biofilm development using a light compound microscope at $40x$ magnification (Microscope Axio Vert.A1 TL/RL, Carl Zeiss AG, ZEISS, Germany).

\subsection{Image analysis and statistics}

Fluorescence images were quantified by measuring the intensity of emission inside the droplets. The average intensity of each droplet was calculated, and each wavelength channel was normalized to the highest mean intensity across all batches. The normalized intensities were then plotted against antibiotic concentration. The experiment was repeated four times, with over $500$ droplets of each type analyzed in total.

Bright-field images were analyzed using an unsupervised learning approach with a VAE. The latent space representation serves as a tool for mapping the structural patterns within the bright-field images, enabling a detailed characterization of the biofilm structures.

Each image was resized to a fixed resolution of $256\times256$ pixels for bright-field images acquired with the light compound microscope. Droplets were detected as circular features using the Hough Circle Transform, an algorithm designed to identify circular shapes and define the entire area of the droplet. 
The detected droplet was cropped to a reduced radius of $4/5$ of the original to isolate the core region, and a Gaussian-blurred mask was applied for a smooth boundary. Thereafter, the processed images were converted to grayscale. By contrast, the time-lapse bright-field images from the inverted microscope were kept at their original size, focusing solely on detecting circular features using the Hough Circle Transform.

\begin{figure}[h!] 
    \centering
    \captionsetup{aboveskip= 10pt, belowskip=0pt} 
    \vspace{1pt}  
    \includegraphics[scale = 0.6]{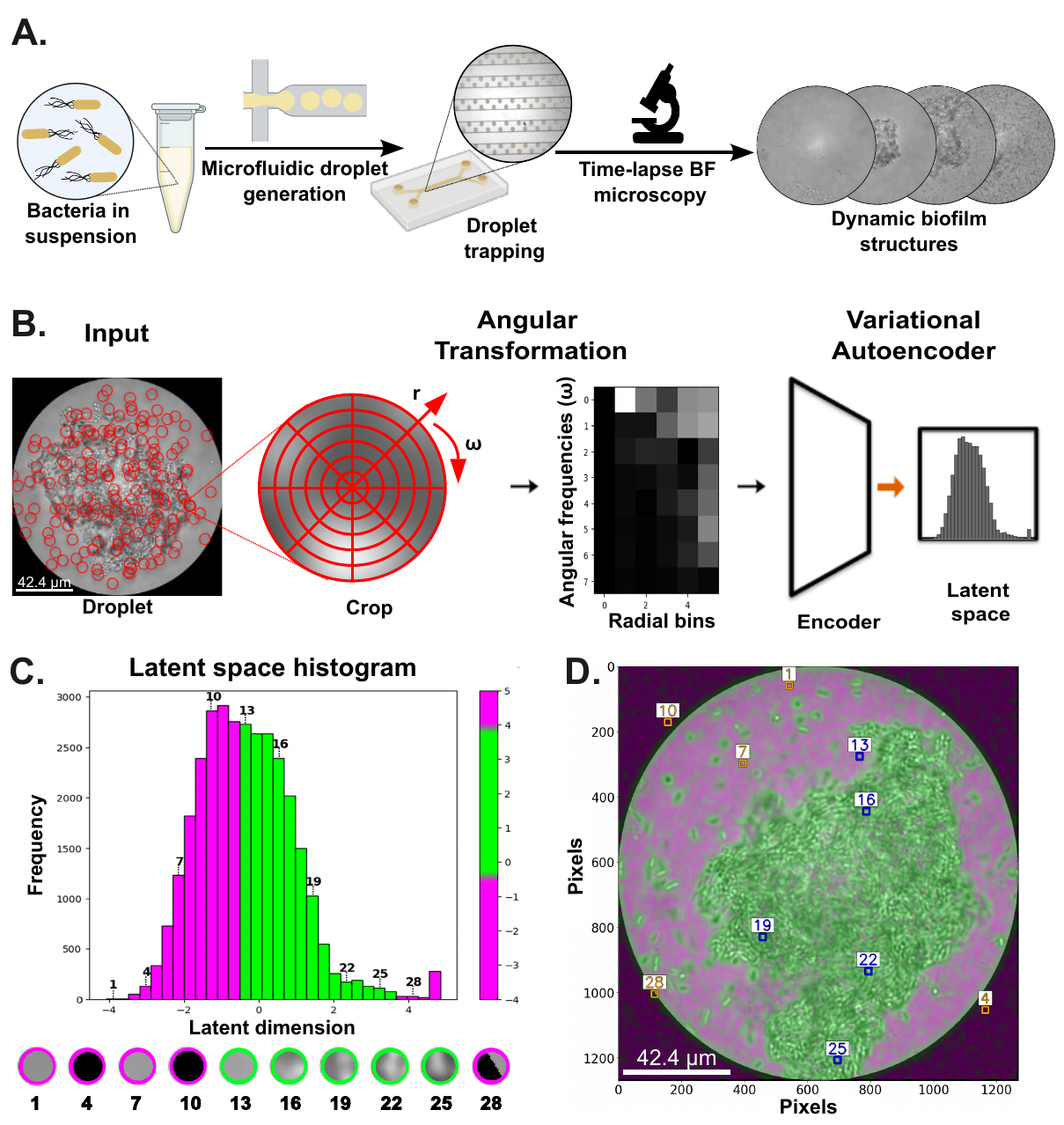}
    \caption{Droplet analysis pipeline and latent space visualization of biofilm images.
    A. In the experimental process, a freshly inoculated bacterial suspension produces and stabilizes droplets inside a microfluidic chip. The development and change of biofilm structures were observed \textit{in situ} using bright-field timelapse microscopy. 
    B. The image processing pipeline includes droplet detection, random cropping, and the angular transformation, which prepare the bright-field images for analysis using the Variational Autoencoder. The processed images serve as the input for feature extraction. 
    C. The histogram represents the latent space of the trained VAE. The chosen color code helps distinguish the droplet background from biofilm structures. Representative elements from different bins along the latent space illustrate the progression from an empty crop to the early formation of biofilm and then to more complex structures.
    D. Visual overlay generated by passing the droplet through the trained VAE using a structured grid. Representative crops from selected histogram bins are overlaid to show their spatial positions within the droplet.}
    \label{fig:fig_1}
\end{figure}

To train the VAE networks, we used two grayscale images per setup: one showing a well-structured biofilm and the other a dispersed sample with individual planktonic bacteria. A separate model was trained for each setup using its corresponding pair of images as input. To enhance spatial diversity and improve pattern recognition, $10,000$ random crops with a defined radius of $12$ pixels were extracted from the training images. These crops were transformed into a radial coordinate system to support rotation-invariant learning, allowing the model to focus on structural features without being affected by the orientation of the input. Specifically, the transformation performs a polar Fourier decomposition, where angular harmonics are analyzed within radial bins. This captures both angular and radial spatial frequency features in a compact form, enabling the VAE to learn meaningful representations of local structures independent of rotation. A visual overview of this preprocessing is shown in Fig.~\ref{fig:fig_1}B. 

The VAE architecture was specifically designed to encode the bright-field image data into a one-dimensional latent representation. The encoder consisted of $1D$ convolutional layers to capture hierarchical features, followed by dense layers that produce the mean and variance of the latent distribution.  
The model was implemented in PyTorch and Lightning using the Deeplay framework \cite{Midtvedt_Deeplay}. Training minimizes two loss components: (1) the reconstruction loss, computed using Binary Cross Entropy Loss (BCELoss), which measures the pixel-wise difference between the original and reconstructed images, and (2) the Kullback-Leibler (KL) divergence, which regularizes the latent space to encourage smoothness and continuity in the learned representations. The model was trained using the Adam optimizer with a learning rate of $0.001$ for $1000$ epochs. Training was performed on an MPS-accelerated backend, utilizing Apple’s Metal Performance Shaders (MPS) for efficient GPU computation.
The training procedure, along with the angular transformation process, is further detailed in SI Section~5.

After training, the encoder extracted one latent variable from each crop, generating compressed representations that captured key image features for downstream mapping and visualization. A histogram of these latent variables was then created to analyze the distribution of features within the learned latent space, offering insights into the underlying data structure. The x-axis corresponded to the single latent dimension, and the y-axis showed the frequency of elements, plotted as a function of this latent space variable (Fig.~\ref{fig:fig_1}C). Each bin was color-coded using a green-magenta colormap, visually corresponding to the different structural patterns in the latent space.

The latent space histogram enabled spatial mapping by identifying which regions of the droplet corresponded to specific latent bins. In this context, magenta represents background regions, including both the interior of the droplet that lacks visible bacterial content and the dark area surrounding the droplet, while green highlights bacterial structures, such as individual cells, aggregates, and larger biofilm formations. These color assignments were determined through visual inspection of the latent space histogram and the associated representative crops. A threshold along the latent dimension was manually selected to separate regions corresponding to background from those containing bacterial content. This choice was guided by a clear visual transition in the crops, from areas showing no bacterial structures to those with distinct biofilm-related features, and was applied consistently across the dataset. As shown in Fig.~\ref{fig:fig_1}D, representative crops are displayed according to their bin positions, illustrating how crops were distributed along the latent dimension. Using this classification, the spatial distribution of biofilm structures within the droplet was mapped, enabling the generation of a visual overlay. To generate this overlay, the mapped colors were blended with the original droplet image to reveal spatial patterns across the field of view. This was done by passing the image through the trained VAE using a structured grid approach that preserves spatial context. Each grid section was assigned a latent value based on its content, which was then used to retrieve the corresponding classification and apply the appropriate color to that region. A stride-based sampling strategy ensured smooth transitions between adjacent areas, minimizing abrupt changes and enhancing spatial continuity. 

This approach laid the foundation for a segmentation pipeline that generated consistent masks to isolate and analyze biofilm structures. These masks enabled the extraction of morphological metrics, such as size and distribution, providing a quantitative basis for understanding the organization of biofilm components within the droplet. The following sections demonstrate how these quantitative metrics provide valuable insights into biofilm morphology. 

Since the data was acquired using two different microscopy setups, each producing datasets with different illumination conditions and imaging parameters, separate models were trained to account for the variability introduced by each optical setup.

\section{Results and discussion}

\subsection{Fluorescence microscopy: Antibiotic sensitivity screening}

One key advantage of the developed platform is its flexibility in adapting to different acquisition modalities. Fluorescence images are highly specific and easy to quantify, whereas bright-field images offer a wider range of information but require advanced techniques for quantification. Therefore, fluorescence is well suited for high-throughput information where statistical relevance is key. 

Biofilm formation under varying antibiotic concentrations was assessed using the genetically modified fluorescent strain MTC871 and the high-throughput droplet approach. After $12$ hours of incubation, a single time point was imaged using fluorescence microscopy to compare the relative abundance of different life stages of \textit{Bacillus subtilis}: motile bacteria and endospores. Endospores, which are highly tolerant and a product of biofilm formation, stand in contrast to motile bacteria, which remain highly susceptible to antibiotics, even at low concentrations. 

\begin{figure}[hbt!] 
    \centering
    \captionsetup{aboveskip = 0pt, belowskip=0pt}  
    \includegraphics[width=\linewidth]{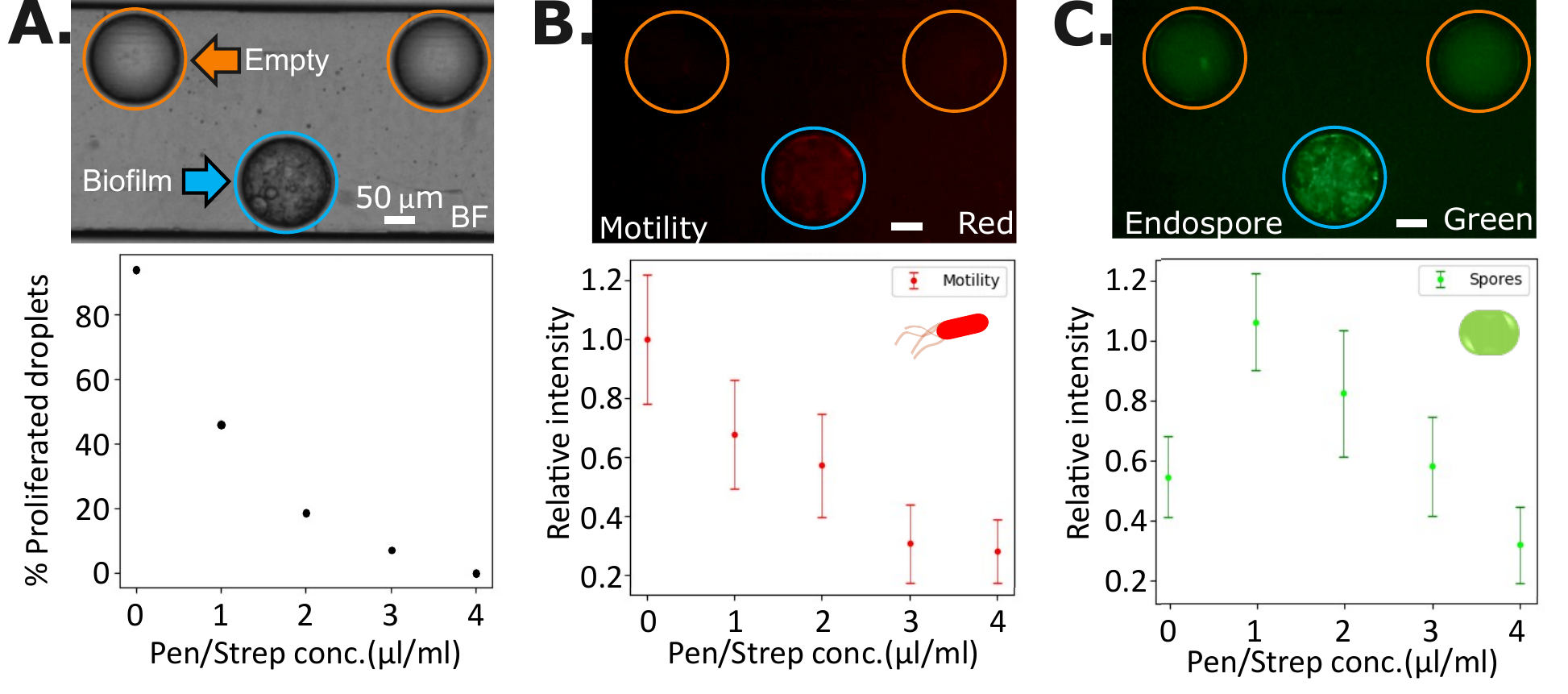}
    \caption{Biofilm formation in the presence of antibiotics. A. Droplets are considered empty or proliferated, and their proportion depends on the presence of antibiotics in the medium.. Increasing antibiotic concentrations reduced the proportion of droplets showing any signs of growth. B. The signal intensity of the red channel indicates the presence of planktonic bacteria. The signal is negligible in empty droplets but varies in proliferated droplets, depending on the presence of antibiotics.  The abundance of planktonic bacteria in proliferated droplets was inversely correlated with antibiotic concentration. C The signal intensity of the green channel indicates the presence of endospores, a consequence of biofilm formation. Empty droplets show a weak signal, while proliferated droplets differ according to antibiotic presence. The highest spore production happened for low concentrations of antibiotic, while in its absence, spore production was comparable to the MIC at $4~\mu L/mL$.}
    \label{fig:fig_2}
\end{figure}

Droplets were imaged at $4\times$ magnification, including multiple droplets per frame, which were then segmented and analyzed individually. Consistent fluorescence intensity and exposure settings were maintained across all antibiotic concentrations for both emission channels. Droplets lacking signal in any channels were classified as empty, indicating an absence of bacterial proliferation and biofilm formation, unlike those where biofilms were observed (see Fig.~\ref{fig:fig_2}A). For each droplet, the mean fluorescence intensity per channel was calculated and subsequently averaged across all biofilm-positive droplets within the same antibiotic concentration group. Final data were normalized relative to the concentration exhibiting the highest intensity for each channel. 

 Increased antibiotic concentration led to decreased biofilm-forming droplets within each batch, as shown in Fig.~\ref{fig:fig_2}A. Given that all droplets were initially populated, the absence of biofilm formation indicated that the antibiotic action had eliminated the initially encapsulated CFU. Initially, the bacteria in suspension were motile and, therefore, more susceptible to antibiotics. Consequently, in droplets where proliferation occurred, the fluorescence signal from motile bacteria diminished with increasing antibiotic concentration (Fig.~\ref{fig:fig_2}B). Interestingly, at sub-inhibitory concentrations well below the Minimum Inhibitory Concentration (MIC) for Pen/Strep ($5$ $\mu L/mL$), biofilm formation was enhanced compared to antibiotics-free conditions (Fig.~\ref{fig:fig_2}C). The fluorescence intensity in these droplets remained comparable to the antibiotic-free case up to $3$ $\mu L/mL$, indicating a broad range of concentrations where antibiotic exposure paradoxically stimulates biofilm development in surviving populations. Beyond this threshold, population survival and biofilm formation were significantly impaired. Notably, the fluorescence intensity signal differed between the two fluorophores, precluding direct quantitative comparisons across channels.
 
 These results reflect a coordinated community response to perceived environmental stress, i.e., antibiotic exposures, as a cue for organization and biofilm formation. They underscore the importance of precise antibiotic use, as inappropriate exposure (particularly at low or trace levels) may inadvertently enhance bacterial tolerance by promoting the formation of biofilms. This highlights the need to consider biofilm-related parameters—such as the concentration window that promotes biofilm formation or the detection threshold for its onset—when evaluating antibiotic efficacy and associated risks.
 
\subsection{Quantitative Analysis of Biofilm Development and Structural Evolution in Droplets}

Bright-field time-lapse imaging provides quantitative and qualitative insights into the dynamic formation of biofilm. Systematic analysis of biofilm growth and structural evolution over time requires quantitative measurements from droplet images. Manual annotations, while possible, are labor-intensive and prone to inconsistency. To address these limitations and accelerate analysis, we employed an unsupervised VAE, enabling rapid and consistent feature extraction.

Due to the constraints of the imaging system, a trade-off was necessary between droplet generation throughput and temporal resolution. The time interval between images was adjusted based on experimental priorities. Higher droplet counts improved the statistical reliability but, in turn, required longer acquisition intervals. Conversely, higher temporal resolution offers more detailed insights into the biofilm life cycle inside droplets. Accordingly, higher throughput was favored for substance effect screening, while higher temporal resolution was prioritized for life cycle studies.

\begin{figure}
    \centering
    \includegraphics[scale=0.49]{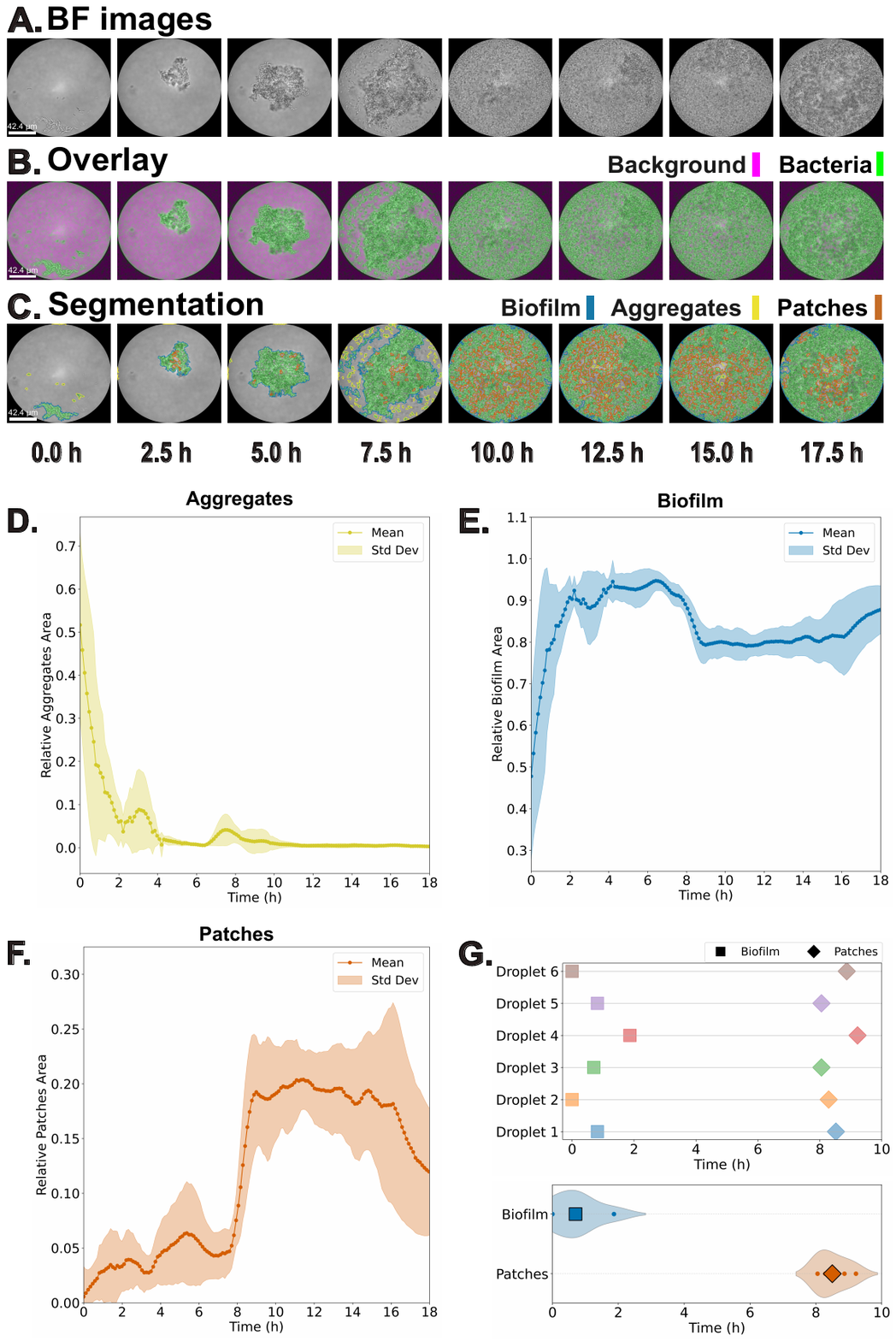}
    \caption{Spatiotemporal analysis of biofilm formation. A. Bright-field time-lapse images show biofilm growth over 17.5 hours. B. Overlay images derived from latent space distribution; green areas denote bacterial presence. C Structural segmentation highlighting spatial distribution: biofilm regions outlined in blue, aggregates in yellow, and patches in orange. D. Relative area of aggregates over time, showing their evolution. E. Temporal changes in relative biofilm area, illustrating dynamic growth patterns. F. Relative area of patches, capturing the emergence of patch-like structures within the biofilm. G. Top: Mean of the positive maximum rate of change for biofilm and patches, summarizing their structural evolution. Bottom: violin plot of the distribution of maximum growth rate for biofilm and patches across six droplets, illustrating variability in their expansion and dispersal. Central symbols indicate mean values. 
    }
    \label{fig:fig_3}
\end{figure}

A time-lapse sequence covering $17.5$ hours with $165$ images (Fig.~\ref{fig:fig_3}A) was processed using the trained VAE and a structured grid approach. This generated latent space feature representations and enabled the generation of overlay images (Fig.~\ref{fig:fig_3}B), where green indicates bacterial structures and biofilms. Magenta represents the droplet background, clearly distinguishing bacterial regions from the surrounding medium.

From these overlays, green masks were used to quantify biofilm growth (Fig.~\ref{fig:fig_3}C). Based on pixel area thresholds, structural components were categorized as Aggregates, Biofilms, or Patches within each droplet image. The contours of aggregates were highlighted in yellow, biofilms in blue, and patches in orange.

Motile, free-swimming bacteria initially formed clusters, distributed throughout the droplet, which we refer to as aggregates. These structures grew over time, gradually losing motility and transitioning into biofilms. Biofilm identification was based not only on size but also on the presence of extracellular matrix (ECM) and adhesion to the droplet interface. For classification, aggregates were defined as bacterial clusters between $300$ and $4999$ pixels$^2$ ($3.32–55.4$ $\mu m^2$), while clusters with an area equal to or greater than $5000$ pixels$^2$ (approximately $55.4$ $\mu m^2$) were classified as biofilm. These biofilm structures often expanded to cover much of the droplet's bottom surface, although growth was not always uniform. Over time, regions with reduced bacterial density—termed patches—emerged within the biofilm. Patches, defined as holes within the biofilm, were retained in the analysis only if their area exceeded $500$ pixels$^2$ ($5.54$ $\mu m^2$) and localized within a biofilm, distinguishing low-density areas from cohesive biofilm. Very small regions below $300$ pixels$^2$ (less than $3.32$ $\mu m^2$) were considered negligible and excluded from the analysis.

After $7.5$ hours, a dispersal phase became evident \cite{Guilhen2017, Kovacs2021, Bartolini2019, Nishikawa2021a}, marked by bacterial detachment from the biofilm structure and a gradual fragmentation of the structure. At later stages, droplets contained a mixture of biofilm, aggregates, patches, and potential residual debris. This complex mixture made differentiating intact biofilm from disintegrating fragments challenging, indicating a transition to a more dispersed state. To characterize this progression, the area occupied by each structural feature was measured relative to the total detected bacteria-containing area at each point, providing insight into biofilm formation, maturation, and dispersal.
Fig.~\ref{fig:fig_3}D illustrates the presence of aggregates, the predominant structures observed during the early stages of the experiment. These aggregates originated from the initial suspension of free-swimming bacteria. Over the first $2$ hours, the relative area occupied by these structures gradually decreased as they progressively merged into larger clusters. Due to the predefined aggregate area threshold, these larger structures were classified as biofilm.
As the remaining motile aggregates increased in size, bacterial motility diminished, causing the aggregates to settle at the base of the droplet (in the imaging focal plane), leading to a temporary increase in aggregate detection between $2$ and $4$ hours. This was followed by their integration into biofilms, resulting in a sustained low aggregate ratio until the $7.5$ hours. At this point, a second peak in aggregate presence occurred, triggered by the rapid escape of bacteria during biofilm dispersal. After approximately $2$ hours, the aggregate ratio dropped again and remained negligible for the rest of the experiment.

Initially, biofilms formed through the merging of aggregates and exhibited rapid growth, as shown in Fig.~\ref{fig:fig_3}E. Over time, biofilms became the dominant structure within the droplet. Although biofilm growth continued steadily through bacterial division, the relative ratios remained stable until the onset of dispersal. An exception to this stability occurred with the sudden appearance of patches caused by the extension of elongated structures that reattached to the biofilm at another point, forming loops. These loops temporarily increased the patch area but were eventually filled in as bacterial division progressed. Once dispersal began, the biofilm ratio declined significantly while the presence of aggregates and patches increased. Following dispersal, some biofilm clusters persisted as stable, coherent structures. A few of these appeared to grow further \cite{ Zhao2017, Nishikawa2021a}, as reflected by the rise in biofilm ratio during the final $2$ hours. These appear as darker regions at the bottom of the droplet that maintain their shape while growing into bigger structures. This phenomenon could be hypothesized to be another biofilm formation cycle after dispersal, where endospores are produced due to nutritional depletion. This correlates with the established relationship between biofilm formation-dispersal and nutrient availability \cite{Gjermansen2005, Harmsen2010, Zhang2014, Hobley2014, Kim2016} and their role in sporulation \cite{Bartolini2019, Kovacs2021}. However, determining whether these structures represent true biofilms or merely accumulating debris would require additional analyses to clarify their chemical and cellular composition. 

Patch evolution is shown in Fig.~\ref{fig:fig_3}F. Initially insignificant, patches appeared during bacterial aggregation and early biofilm formation due to random cell orientation. As \textit{Bacillus subtilis} divides, it often forms aligned chains, which occasionally loop and reattach, as seen in Fig.~\ref{fig:fig_4}A at the $3$h $20'$ timepoint, temporarily increasing patch ratio area (shown as the first relative maximum of the curve). These loops contributed to the first peak in patch ratio and the corresponding increase in standard deviation, driven by stochastic loop formation and size variability. As loops were filled in, the patch ratio declined until dispersal, where a sharp increase signaled structure breakdown. This stage occurred consistently across droplets and was marked by a sustained rise in patch presence, which gradually declined towards the end of the experiment. 

The dynamics of aggregates, biofilm, and patches collectively offer a comprehensive view of biofilm lifecycle events in nutrient-limited spaces. Two key transitions were identified: the biofilm formation trigger, defined as the point of maximum growth rate of the biofilm (i.e., the peak of the derivative of the biofilm area curve), and the dispersal trigger, defined as the point of maximum growth rate of patch formation. To capture the variability across droplets, we analyzed the timing and distribution of these transitions in all samples. Fig.~\ref{fig:fig_3}G displays six recorded droplets, showing the temporal evolution of biofilm structures. Squares and diamonds indicate the time points of peak positive growth for biofilm and patches, respectively, as determined by the derivatives of their relative area ratio curves. Violin plots reveal the distribution of these events, highlighting variability and trends in biofilm and patch formation across the different droplets.

These findings, supported by the unsupervised model that generates feature-based overlays, offer a data-driven perspective on droplet biofilm dynamics. The minimal required training demonstrates the robustness of the VAE model in generalizing feature representation from only two distinctly different microscopy images. Biofilm, aggregates and patches can be classified through model-based segmentation, uncovering a dynamic cycle of growth and dispersal, as evidenced by sequential peaks in their area ratios. This method not only aligns with visual observations but also highlights the potential of unsupervised approaches for reliable, high-throughput analysis of complex biological structures, as further demonstrated in the following section.

\subsection{Quantitative Analysis of Biofilm Modifiers}

After establishing a baseline for analyzing biofilm dynamics from bright-field time-lapse microscopy images, we applied the workflow to a relevant case study: assessing the impact of biofilm-modifying substances. For this experiment, three types of droplets were generated within the same microfluidic chip: (i) sterile Lysogeny Broth medium serving as a contamination control, (ii) inoculated Lysogeny Broth medium as a negative control, and (iii) inoculated Lysogeny Broth medium supplemented with $1\%$ (v/v) glycerol as the biofilm-promoting agent for \textit{Bacillus subtilis}, making it an ideal candidate to validate the analysis pipeline. Images of the bacterial populations within droplets were acquired and analyzed based on the presence or absence of glycerol. In this case, the focus was on throughput rather than temporal resolution. Accordingly, batches of over $50$ droplets per condition were imaged at $40$ minute intervals over an $8$ hour observation period, covering the full biofilm development cycle until dispersal begins. After dispersal, potential biofilm segmentation became increasingly challenging for the current model due to the presence of swimming bacteria and debris, so we focused exclusively on the formation process. The resulting images were processed using the previously described unsupervised model.
 
\begin{figure}[h!]
    \centering
    \includegraphics[scale = 0.5]{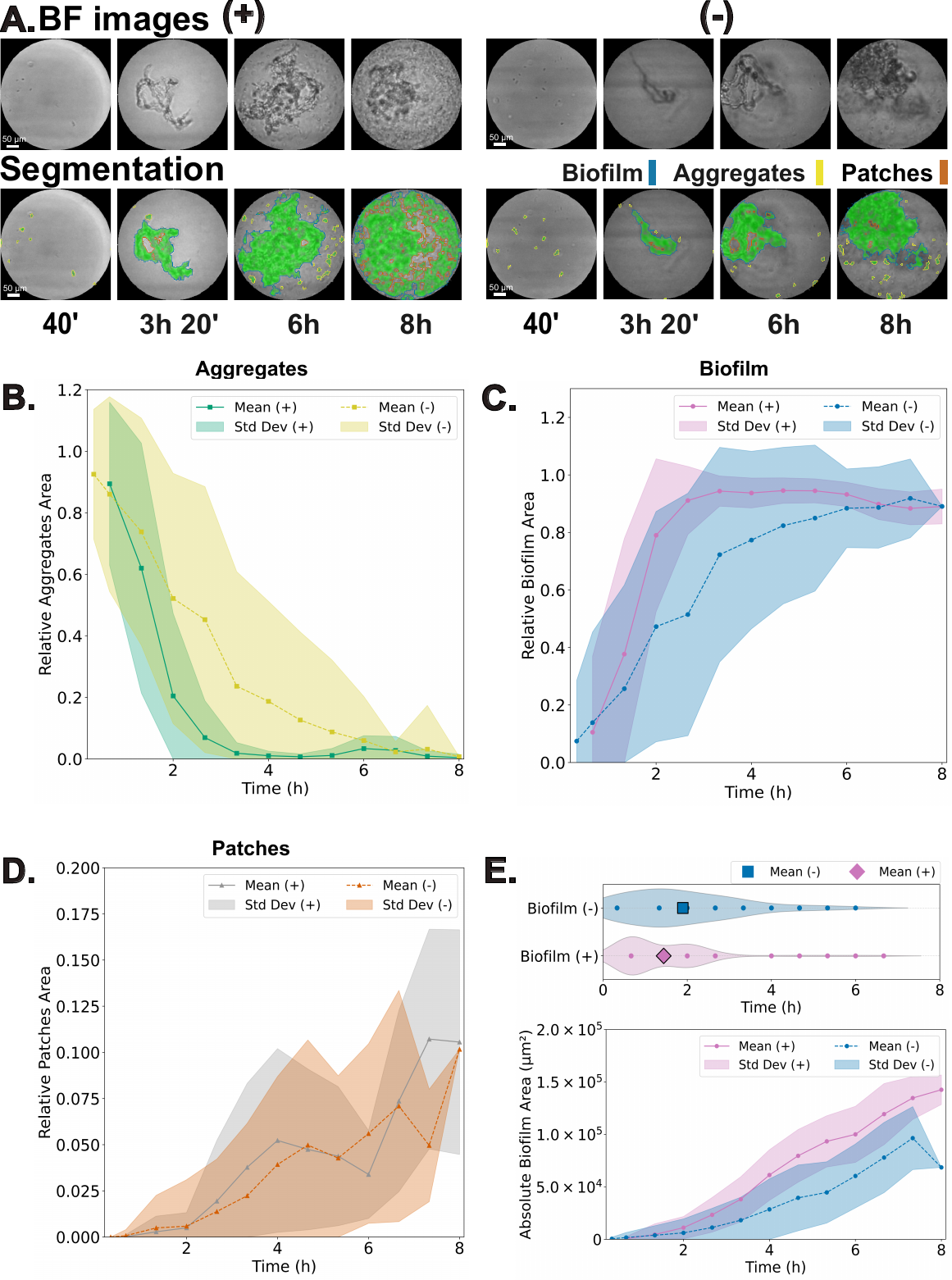}
    \caption{Influence of glycerol on biofilm formation A. Representative images of biofilm formation at $40$ minutes, $3$ hours, $20$ minutes, $6$ hours, and $8$ hours, under conditions with (+) and without (-) glycerol. B. Relative aggregate area, showing the evolution of aggregates with and without glycerol. C. Relative biofilm area over time, illustrating differences in biofilm growth dynamics between the two conditions. D. Relative patch area, indicating the formation of patches within the biofilm structure under both conditions. E. Violin plot of maximum biofilm growth rate across multiple droplets, illustrating the distribution of growth dynamics under both conditions. This quantitative analysis provides insights into biofilm formation's temporal evolution and glycerol's influence on its structural development.}
    \label{fig:fig_4}
\end{figure}

Sterile Lysogeny Broth droplets showed no signs of bacterial proliferation, confirming the absence of contamination and validating the platform's reliability. Qualitative analysis revealed that droplets containing glycerol exhibited faster biofilm formation and showed early signs of dispersal as soon as $6$ hours into the experiment Fig.~\ref{fig:fig_4}A. In contrast, droplets without glycerol developed more slowly, mirroring the dynamics observed in the time-resolved experiments. Still, the biofilm structures described were consistent among both types of droplets.

Quantitatively, the most notable distinction emerged in the statistical distribution of the data. Droplets containing glycerol exhibited a narrower distribution and lower variability, suggesting that glycerol may act more as a signaling molecule than as supplementary nutrition. Specifically, aggregate formation occurred over a shorter time window in the presence of glycerol, and a local maximum in the ratio appeared at $6$ hours—likely marking the onset of dispersal (Fig.~\ref{fig:fig_4}B). In contrast, droplets without glycerol only showed a mild increase in aggregate ratio after $7$ hours. The biofilm ratio (Fig.~\ref{fig:fig_4}C) followed a similar trend. In droplets with glycerol, biofilms formed more rapidly and reached a stable level within $3$ hours. In the absence of glycerol, a plateau was only reached after $4$ hours, consistent with the control data from the previous result section. Patches formation (Fig.~\ref{fig:fig_4}D) followed a comparable progression in both conditions up to $6$ hours. While glycerol appeared to slightly accelerate early development, it did not significantly affect the formation of the looped structures that define patches at this stage. However, after $6$ hours, droplets with glycerol showed a sharp increase in the patch ratio –signaling active dispersal— while droplets without glycerol exhibited this pattern only at the final time point. As a summary metric, the maximum rate of biofilm formation (i.e., the point of fastest biofilm growth) was calculated for both conditions and shown in Fig.~\ref{fig:fig_4}E. This metric reflects the overall trends in the time series: glycerol induced earlier and more synchronized biofilm formation, while in its absence, the process was slower and more variable. 

An absolute representation of biofilm area over time further supported these findings, showing that droplets with glycerol consistently reached larger biofilm-covered areas. Interestingly, although the absolute standard deviation of biofilm sizes was similar between both conditions, their relative variability (as seen in the ratio-based analysis) differed. This highlights the importance of normalized, comparative metrics in identifying patterns within inherently variable biological systems. Moreover, it reinforces the idea that glycerol influences not only biofilm mass but also bacterial organization. Overall, these results validate the platform's effectiveness in quantifying the influence of biofilm modifiers and demonstrate the potential to uncover new insights into the mechanisms driving biofilm regulation.

\section{Conclusions}
We have developed a robust and versatile platform for both qualitative and quantitative studies of biofilms within microfluidic droplets. This platform has been successfully applied to various use cases, including antibiotic sensitivity screening, exploration of biofilm life cycle dynamics, and evaluation of a biofilm-promoting substance. Its flexibility has been demonstrated across different imaging modalities,  analysis methods, and degrees of automation.

A core strength of the platform is its integration of an unsupervised model for automated segmentation and quantification. Using a VAE, high-dimensional image data is encoded into a lower-dimensional latent space that retains essential structural information. This provides a powerful framework for longitudinal studies, enabling consistent and precise monitoring and tracking of biofilm formation and structural transitions over time.

This VAE-based segmentation enables autonomous classification and measurement of biofilm, aggregates, and patch areas. By generating masks from overlay images and applying thresholding techniques, biofilm structures are accurately isolated and quantified. The observed sequential peaks in biofilm, aggregate, and patch area ratios offer detailed insights into the biofilm life cycle, revealing structured growth and dispersal phases. This quantitative output is key to advancing our understanding of biofilm morphology and dynamics within confined microenvironments, including dispersal and potential re-formation phases. 

Further enhancement of the model—such as expanding the dimensions of the latent space—could improve its ability to identify additional features, including cell alignment, endospore formation, and microbial diversity. This opens new possibilities for detecting various biofilm components, bacterial phenotypes, and even different species involved in the biofilm matrix.

Overall, incorporating VAEs into the analysis pipeline significantly deepens our ability to study biofilms in detail. Coupled with the versatile droplet-based microfluidic system and various imaging and analysis options, this platform offers a comprehensive and scalable solution for future investigations into biofilm formation, growth, dispersal, and re-growth. It paves the way for new insights and innovations in fundamental research and applied contexts addressing biofilm-related challenges.

\section*{Funding and acknowledgments}
This project received support from the European Union's Horizon 2020 research and innovation programme within the Active Matter ITN (grant No. 812780), the Swedish Research Council (grant No. 2019-05401), the Horizon Europe ERC Consolidator Grant MAPEI (grant No. 101001267) and the Knut and Alice Wallenberg Foundation (grant No 2019.0079). Jesus Manuel Antúnez Domínguez thanks the Adlerbertska forskningsstiftelsen program from 2023 and 2024. Thanks also to Dr. Matthew Cabeen, Assistant Professor in the Department of Microbiology and Molecular Genetics of Oklahoma State University, who kindly provided the strain of \textit{Bacillus subtilis} used in this project.

\section*{Author Contributions}

Daniela Pérez Guerrero: 
Conceptualization, data curation, formal analysis, funding acquisition, methodology, software, visualization, writing – original draft.
\newline
Jesus Manuel Antúnez Domínguez:
Conceptualization, data curation, funding acquisition, methodology, investigation, visualization, writing – original draft.
\newline
Aurélie Vigne:
Conceptualization, project administration, resources, supervision, writing – review \& editing.
\newline
Daniel Midtvedt:
Methodology, software, writing – review \& editing.
\newline
Wylie Ahmed:
Conceptualization, methodology, software, supervision, writing – review \& editing.
\newline 
Lisa D. Muiznieks:
Conceptualization, project administration, resources, supervision, writing – review \& editing.
\newline
Giovanni Volpe:
Conceptualization, funding acquisition, methodology, software, supervision, writing – review \& editing.
\newline
Caroline Beck Adiels:
Conceptualization, funding acquisition, methodology, project administration, supervision, resources, visualization, writing – review \& editing.
\newline

\section*{Competing Interests statement}

The authors declare that they have no known competing financial interests or personal relationships that could have appeared to influence the work reported in this paper.

\section*{Data availability}
Data will be made available on request.

\appendix
 \bibliographystyle{elsarticle-num} 
 \bibliography{sample}

\begin{thebibliography}{10}
\expandafter\ifx\csname url\endcsname\relax
  \def\url#1{\texttt{#1}}\fi
\expandafter\ifx\csname urlprefix\endcsname\relax\def\urlprefix{URL }\fi
\expandafter\ifx\csname href\endcsname\relax
  \def\href#1#2{#2} \def\path#1{#1}\fi

\bibitem{Popat2015}
R.~Popat, D.~M. Cornforth, L.~McNally, S.~P. Brown, \href{https://royalsocietypublishing.org/doi/10.1098/rsif.2014.0882}{Collective sensing and collective responses in quorum-sensing bacteria}, Journal of The Royal Society Interface 12 (2015) 20140882.
\newblock \href {https://doi.org/10.1098/rsif.2014.0882} {\path{doi:10.1098/rsif.2014.0882}}.
\newline\urlprefix\url{https://royalsocietypublishing.org/doi/10.1098/rsif.2014.0882}

\bibitem{Mitchell2006}
J.~G. Mitchell, K.~Kogure, \href{http://www.ncbi.nlm.nih.gov/pubmed/16420610}{Bacterial motility: links to the environment and a driving force for microbial physics.}, FEMS microbiology ecology 55 (2006) 3--16.
\newblock \href {https://doi.org/10.1111/j.1574-6941.2005.00003.x} {\path{doi:10.1111/j.1574-6941.2005.00003.x}}.
\newline\urlprefix\url{http://www.ncbi.nlm.nih.gov/pubmed/16420610}

\bibitem{Wadhwa2022}
N.~Wadhwa, H.~C. Berg, \href{http://www.ncbi.nlm.nih.gov/pubmed/34548639}{Bacterial motility: machinery and mechanisms.}, Nature reviews. Microbiology 20 (2022) 161--173.
\newblock \href {https://doi.org/10.1038/s41579-021-00626-4} {\path{doi:10.1038/s41579-021-00626-4}}.
\newline\urlprefix\url{http://www.ncbi.nlm.nih.gov/pubmed/34548639}

\bibitem{Miller2001}
M.~B. Miller, B.~L. Bassler, \href{https://www.annualreviews.org/doi/10.1146/annurev.micro.55.1.165}{Quorum sensing in bacteria}, Annual Review of Microbiology 55 (2001) 165--199.
\newblock \href {https://doi.org/10.1146/annurev.micro.55.1.165} {\path{doi:10.1146/annurev.micro.55.1.165}}.
\newline\urlprefix\url{https://www.annualreviews.org/doi/10.1146/annurev.micro.55.1.165}

\bibitem{Parsek2005}
M.~R. Parsek, E.~Greenberg, \href{https://linkinghub.elsevier.com/retrieve/pii/S0966842X04002616}{Sociomicrobiology: the connections between quorum sensing and biofilms}, Trends in Microbiology 13 (2005) 27--33.
\newblock \href {https://doi.org/10.1016/j.tim.2004.11.007} {\path{doi:10.1016/j.tim.2004.11.007}}.
\newline\urlprefix\url{https://linkinghub.elsevier.com/retrieve/pii/S0966842X04002616}

\bibitem{Arnaouteli2021}
S.~Arnaouteli, N.~C. Bamford, N.~R. Stanley-Wall, Ákos T.~Kovács, \href{https://www.nature.com/articles/s41579-021-00540-9}{Bacillus subtilis biofilm formation and social interactions}, Nature Reviews Microbiology 19 (2021) 600--614.
\newblock \href {https://doi.org/10.1038/s41579-021-00540-9} {\path{doi:10.1038/s41579-021-00540-9}}.
\newline\urlprefix\url{https://www.nature.com/articles/s41579-021-00540-9}

\bibitem{Wucher2021}
B.~R. Wucher, M.~Elsayed, J.~S. Adelman, D.~E. Kadouri, C.~D. Nadell, Bacterial predation transforms the landscape and community assembly of biofilms, Current Biology 31 (2021) 2643--2651.e3.
\newblock \href {https://doi.org/10.1016/j.cub.2021.03.036} {\path{doi:10.1016/j.cub.2021.03.036}}.

\bibitem{Espinal2012}
P.~Espinal, S.~Martí, J.~Vila, \href{https://linkinghub.elsevier.com/retrieve/pii/S0195670111003458}{Effect of biofilm formation on the survival of acinetobacter baumannii on dry surfaces}, Journal of Hospital Infection 80 (2012) 56--60.
\newblock \href {https://doi.org/10.1016/j.jhin.2011.08.013} {\path{doi:10.1016/j.jhin.2011.08.013}}.
\newline\urlprefix\url{https://linkinghub.elsevier.com/retrieve/pii/S0195670111003458}

\bibitem{Hathroubi2017}
S.~Hathroubi, M.~A. Mekni, P.~Domenico, D.~Nguyen, M.~Jacques, \href{http://www.liebertpub.com/doi/10.1089/mdr.2016.0087}{Biofilms: Microbial shelters against antibiotics}, Microbial Drug Resistance 23 (2017) 147--156.
\newblock \href {https://doi.org/10.1089/mdr.2016.0087} {\path{doi:10.1089/mdr.2016.0087}}.
\newline\urlprefix\url{http://www.liebertpub.com/doi/10.1089/mdr.2016.0087}

\bibitem{Perez2020}
J.~Pérez, F.~J. Contreras-Moreno, F.~J. Marcos-Torres, A.~Moraleda-Muñoz, J.~Muñoz-Dorado, \href{https://linkinghub.elsevier.com/retrieve/pii/S2001037020303949}{The antibiotic crisis: How bacterial predators can help}, Computational and Structural Biotechnology Journal 18 (2020) 2547--2555.
\newblock \href {https://doi.org/10.1016/j.csbj.2020.09.010} {\path{doi:10.1016/j.csbj.2020.09.010}}.
\newline\urlprefix\url{https://linkinghub.elsevier.com/retrieve/pii/S2001037020303949}

\bibitem{Sun2019}
D.~Sun, K.~Jeannot, Y.~Xiao, C.~W. Knapp, \href{https://www.frontiersin.org/article/10.3389/fmicb.2019.01933/full}{Editorial: Horizontal gene transfer mediated bacterial antibiotic resistance}, Frontiers in Microbiology 10 (8 2019).
\newblock \href {https://doi.org/10.3389/fmicb.2019.01933} {\path{doi:10.3389/fmicb.2019.01933}}.
\newline\urlprefix\url{https://www.frontiersin.org/article/10.3389/fmicb.2019.01933/full}

\bibitem{Kolpen2022}
M.~Kolpen, K.~N. Kragh, J.~B. Enciso, D.~Faurholt-Jepsen, B.~Lindegaard, G.~B. Egelund, A.~V. Jensen, P.~Ravn, I.~H.~M. Mathiesen, A.~G. Gheorge, F.~B. Hertz, T.~Qvist, M.~Whiteley, P.~Østrup Jensen, T.~Bjarnsholt, Bacterial biofilms predominate in both acute and chronic human lung infections, Thorax 77 (2022) 1015--1022.
\newblock \href {https://doi.org/10.1136/thoraxjnl-2021-217576} {\path{doi:10.1136/thoraxjnl-2021-217576}}.

\bibitem{Sonawane2022}
J.~M. Sonawane, A.~K. Rai, M.~Sharma, M.~Tripathi, R.~Prasad, Microbial biofilms: Recent advances and progress in environmental bioremediation, Science of the Total Environment 824 (6 2022).
\newblock \href {https://doi.org/10.1016/j.scitotenv.2022.153843} {\path{doi:10.1016/j.scitotenv.2022.153843}}.

\bibitem{Linton1999}
C.~J. Linton, A.~Sherriff, M.~R. Millar, \href{https://academic.oup.com/jambio/article/86/2/194/6720783}{Use of a modified robbins device to directly compare the adhesion of staphylococcus epidermidis rp62a to surfaces}, Journal of Applied Microbiology 86 (1999) 194--202.
\newblock \href {https://doi.org/10.1046/j.1365-2672.1999.00650.x} {\path{doi:10.1046/j.1365-2672.1999.00650.x}}.
\newline\urlprefix\url{https://academic.oup.com/jambio/article/86/2/194/6720783}

\bibitem{Schwartz2010}
K.~Schwartz, R.~Stephenson, M.~Hernandez, N.~Jambang, B.~Boles, \href{https://app.jove.com/t/2470}{The use of drip flow and rotating disk reactors for <em>staphylococcus aureus</em> biofilm analysis}, Journal of Visualized Experiments (12 2010).
\newblock \href {https://doi.org/10.3791/2470} {\path{doi:10.3791/2470}}.
\newline\urlprefix\url{https://app.jove.com/t/2470}

\bibitem{Azeredo2017}
J.~Azeredo, N.~F. Azevedo, R.~Briandet, N.~Cerca, T.~Coenye, A.~R. Costa, M.~Desvaux, G.~D. Bonaventura, M.~Hébraud, Z.~Jaglic, M.~Kačániová, S.~Knøchel, A.~Lourenço, F.~Mergulhão, R.~L. Meyer, G.~Nychas, M.~Simões, O.~Tresse, C.~Sternberg, \href{https://www.tandfonline.com/doi/full/10.1080/1040841X.2016.1208146}{Critical review on biofilm methods}, Critical Reviews in Microbiology 43 (2017) 313--351.
\newblock \href {https://doi.org/10.1080/1040841X.2016.1208146} {\path{doi:10.1080/1040841X.2016.1208146}}.
\newline\urlprefix\url{https://www.tandfonline.com/doi/full/10.1080/1040841X.2016.1208146}

\bibitem{Kaminski2016a}
T.~S. Kaminski, O.~Scheler, P.~Garstecki, \href{http://dx.doi.org/10.1039/C6LC00367B http://xlink.rsc.org/?DOI=C6LC00367B}{Droplet microfluidics for microbiology: techniques, applications and challenges}, Lab on a Chip 16 (2016) 2168--2187.
\newblock \href {https://doi.org/10.1039/C6LC00367B} {\path{doi:10.1039/C6LC00367B}}.
\newline\urlprefix\url{http://dx.doi.org/10.1039/C6LC00367B http://xlink.rsc.org/?DOI=C6LC00367B}

\bibitem{Niculescu2021}
A.~G. Niculescu, C.~Chircov, A.~C. Bîrcă, A.~M. Grumezescu, Fabrication and applications of microfluidic devices: A review (2 2021).
\newblock \href {https://doi.org/10.3390/ijms22042011} {\path{doi:10.3390/ijms22042011}}.

\bibitem{Chen2022}
Z.~Chen, S.~Kheiri, E.~W.~K. Young, E.~Kumacheva, \href{https://pubs.acs.org/doi/10.1021/acs.langmuir.2c00491}{Trends in droplet microfluidics: From droplet generation to biomedical applications}, Langmuir 38 (2022) 6233--6248.
\newblock \href {https://doi.org/10.1021/acs.langmuir.2c00491} {\path{doi:10.1021/acs.langmuir.2c00491}}.
\newline\urlprefix\url{https://pubs.acs.org/doi/10.1021/acs.langmuir.2c00491}

\bibitem{Jiang2022}
M.-Z. Jiang, H.-Z. Zhu, N.~Zhou, C.~Liu, C.-Y. Jiang, Y.~Wang, S.-J. Liu, \href{https://doi.org/10.1038/s41598-022-23000-7}{Droplet microfluidics-based high-throughput bacterial cultivation for validation of taxon pairs in microbial co-occurrence networks}, Scientific Reports 12~(1) (2022) 18145.
\newblock \href {https://doi.org/10.1038/s41598-022-23000-7} {\path{doi:10.1038/s41598-022-23000-7}}.
\newline\urlprefix\url{https://doi.org/10.1038/s41598-022-23000-7}

\bibitem{kingma2014auto}
D.~P. Kingma, M.~Welling, Auto-encoding variational bayes, in: Proceedings of the 2nd International Conference on Learning Representations (ICLR), 2014, arXiv:1312.6114.

\bibitem{Volpe2025}
G.~Volpe, B.~Midtvedt, J.~Pineda, H.~K. Moberg, H.~Bachimanchi, J.~B. Pereira, C.~Manzo, Deep Learning Crash Course, No Starch Press, 2025.

\bibitem{Gu2019}
H.-J. Gu, Q.-L. Sun, J.-C. Luo, J.~Zhang, L.~Sun, \href{https://www.frontiersin.org/article/10.3389/fcimb.2019.00183/full}{A first study of the virulence potential of a bacillus subtilis isolate from deep-sea hydrothermal vent}, Frontiers in Cellular and Infection Microbiology 9 (5 2019).
\newblock \href {https://doi.org/10.3389/fcimb.2019.00183} {\path{doi:10.3389/fcimb.2019.00183}}.
\newline\urlprefix\url{https://www.frontiersin.org/article/10.3389/fcimb.2019.00183/full}

\bibitem{Liu2023}
H.~Liu, L.~Nan, F.~Chen, Y.~Zhao, Y.~Zhao, \href{http://xlink.rsc.org/?DOI=D3LC00224A}{Functions and applications of artificial intelligence in droplet microfluidics}, Lab on a Chip 23 (2023) 2497--2513.
\newblock \href {https://doi.org/10.1039/D3LC00224A} {\path{doi:10.1039/D3LC00224A}}.
\newline\urlprefix\url{http://xlink.rsc.org/?DOI=D3LC00224A}

\bibitem{Feng2022}
C.~Feng, K.~Takahashi, J.~Zhu, \href{https://www.frontiersin.org/articles/10.3389/fbioe.2022.891213/full}{Simple one-step and rapid patterning of pdms microfluidic device wettability for pdms shell production}, Frontiers in Bioengineering and Biotechnology 10 (4 2022).
\newblock \href {https://doi.org/10.3389/fbioe.2022.891213} {\path{doi:10.3389/fbioe.2022.891213}}.
\newline\urlprefix\url{https://www.frontiersin.org/articles/10.3389/fbioe.2022.891213/full}

\bibitem{Midtvedt_Deeplay}
B.~Midtvedt, J.~Pineda, H.~Klein~Morberg, H.~Bachimanchi, M.~Granfors, A.~Lech, C.~Manzo, G.~Volpe, \href{https://github.com/DeepTrackAI/deeplay}{{Deeplay}}.
\newline\urlprefix\url{https://github.com/DeepTrackAI/deeplay}

\bibitem{Guilhen2017}
C.~Guilhen, C.~Forestier, D.~Balestrino, \href{https://onlinelibrary.wiley.com/doi/10.1111/mmi.13698}{Biofilm dispersal: multiple elaborate strategies for dissemination of bacteria with unique properties}, Molecular Microbiology 105 (2017) 188--210.
\newblock \href {https://doi.org/10.1111/mmi.13698} {\path{doi:10.1111/mmi.13698}}.
\newline\urlprefix\url{https://onlinelibrary.wiley.com/doi/10.1111/mmi.13698}

\bibitem{Kovacs2021}
Ákos T~Kovács, N.~R. Stanley-Wall, \href{http://www.ncbi.nlm.nih.gov/pubmed/33927051 http://www.pubmedcentral.nih.gov/articlerender.fcgi?artid=PMC8223919}{Biofilm dispersal for spore release in bacillus subtilis.}, Journal of bacteriology 203 (2021) e0019221.
\newblock \href {https://doi.org/10.1128/JB.00192-21} {\path{doi:10.1128/JB.00192-21}}.
\newline\urlprefix\url{http://www.ncbi.nlm.nih.gov/pubmed/33927051 http://www.pubmedcentral.nih.gov/articlerender.fcgi?artid=PMC8223919}

\bibitem{Bartolini2019}
M.~Bartolini, S.~Cogliati, D.~Vileta, C.~Bauman, L.~Rateni, C.~Leñini, F.~Argañaraz, M.~Francisco, J.~M. Villalba, L.~Steil, U.~Völker, R.~Grau, \href{https://journals.asm.org/doi/10.1128/JB.00473-18}{Regulation of biofilm aging and dispersal in bacillus subtilis by the alternative sigma factor sigb}, Journal of Bacteriology 201 (1 2019).
\newblock \href {https://doi.org/10.1128/JB.00473-18} {\path{doi:10.1128/JB.00473-18}}.
\newline\urlprefix\url{https://journals.asm.org/doi/10.1128/JB.00473-18}

\bibitem{Nishikawa2021a}
M.~Nishikawa, K.~Kobayashi, \href{https://journals.asm.org/doi/10.1128/JB.00114-21}{Calcium prevents biofilm dispersion in bacillus subtilis}, Journal of Bacteriology 203 (6 2021).
\newblock \href {https://doi.org/10.1128/JB.00114-21} {\path{doi:10.1128/JB.00114-21}}.
\newline\urlprefix\url{https://journals.asm.org/doi/10.1128/JB.00114-21}

\bibitem{Zhao2017}
R.~Zhao, Y.~Song, Q.~Dai, Y.~Kang, J.~Pan, L.~Zhu, L.~Zhang, Y.~Wang, X.~Shen, \href{http://dx.doi.org/10.1038/s41598-017-00534-9 http://www.nature.com/articles/s41598-017-00534-9}{A starvation-induced regulator, rovm, acts as a switch for planktonic/biofilm state transition in yersinia pseudotuberculosis}, Scientific Reports 7 (2017) 639.
\newblock \href {https://doi.org/10.1038/s41598-017-00534-9} {\path{doi:10.1038/s41598-017-00534-9}}.
\newline\urlprefix\url{http://dx.doi.org/10.1038/s41598-017-00534-9 http://www.nature.com/articles/s41598-017-00534-9}

\bibitem{Gjermansen2005}
M.~Gjermansen, P.~Ragas, C.~Sternberg, S.~Molin, T.~Tolker-Nielsen, \href{https://onlinelibrary.wiley.com/doi/10.1111/j.1462-2920.2005.00775.x}{Characterization of starvation-induced dispersion in pseudomonas putida biofilms}, Environmental Microbiology 7 (2005) 894--904.
\newblock \href {https://doi.org/10.1111/j.1462-2920.2005.00775.x} {\path{doi:10.1111/j.1462-2920.2005.00775.x}}.
\newline\urlprefix\url{https://onlinelibrary.wiley.com/doi/10.1111/j.1462-2920.2005.00775.x}

\bibitem{Harmsen2010}
M.~Harmsen, L.~Yang, S.~J. Pamp, T.~Tolker-Nielsen, \href{https://academic.oup.com/femspd/article-lookup/doi/10.1111/j.1574-695X.2010.00690.x}{An update on pseudomonas aeruginosa biofilm formation, tolerance, and dispersal}, FEMS Immunology \& Medical Microbiology 59 (2010) 253--268.
\newblock \href {https://doi.org/10.1111/j.1574-695X.2010.00690.x} {\path{doi:10.1111/j.1574-695X.2010.00690.x}}.
\newline\urlprefix\url{https://academic.oup.com/femspd/article-lookup/doi/10.1111/j.1574-695X.2010.00690.x}

\bibitem{Zhang2014}
W.~Zhang, A.~Seminara, M.~Suaris, M.~P. Brenner, D.~A. Weitz, T.~E. Angelini, \href{https://iopscience.iop.org/article/10.1088/1367-2630/16/1/015028}{Nutrient depletion in bacillus subtilis biofilms triggers matrix production}, New Journal of Physics 16 (2014) 015028.
\newblock \href {https://doi.org/10.1088/1367-2630/16/1/015028} {\path{doi:10.1088/1367-2630/16/1/015028}}.
\newline\urlprefix\url{https://iopscience.iop.org/article/10.1088/1367-2630/16/1/015028}

\bibitem{Hobley2014}
L.~Hobley, S.~Kim, Y.~Maezato, S.~Wyllie, A.~Fairlamb, N.~Stanley-Wall, A.~Michael, \href{https://linkinghub.elsevier.com/retrieve/pii/S0092867414000233}{Norspermidine is not a self-produced trigger for biofilm disassembly}, Cell 156 (2014) 844--854.
\newblock \href {https://doi.org/10.1016/j.cell.2014.01.012} {\path{doi:10.1016/j.cell.2014.01.012}}.
\newline\urlprefix\url{https://linkinghub.elsevier.com/retrieve/pii/S0092867414000233}

\bibitem{Kim2016}
S.-K. Kim, J.-H. Lee, \href{http://link.springer.com/10.1007/s12275-016-5528-7}{Biofilm dispersion in pseudomonas aeruginosa}, Journal of Microbiology 54 (2016) 71--85.
\newblock \href {https://doi.org/10.1007/s12275-016-5528-7} {\path{doi:10.1007/s12275-016-5528-7}}.
\newline\urlprefix\url{http://link.springer.com/10.1007/s12275-016-5528-7}

\end{thebibliography}

\end{document}


\renewcommand{\figurename}{Fig.}
\maketitle

\section{\textit{B. subtilis} growth and calibration curves}
\textit{B. subtilis} strain MTC871 is genetically modified from the wild type NCIB 3610, adding  modifications to express fluorescence:

\begin{enumerate}

  \item sacA::Phag-mkate2 [KanR] Fluorescent red (Fluorophore mKate) when motile (flagella hag gene expressed in the bacteria)

  \item amyE::PtapA-cfp [SpcR]  Fluorescent blue (Fluorophore CFP) when sessile (matrix production operon tapA-sipW-tasA expressed in the bacteria)

  \item ywrK::PsspB-citrus [CmR]) Fluorescent green (Fluorophore citrus) when dormant (endospore formation sspB gene activated)

\end{enumerate}

\label{strain}

\begin{figure}[H]
\begin{center}
\resizebox{0.95\columnwidth}{!}
{\includegraphics{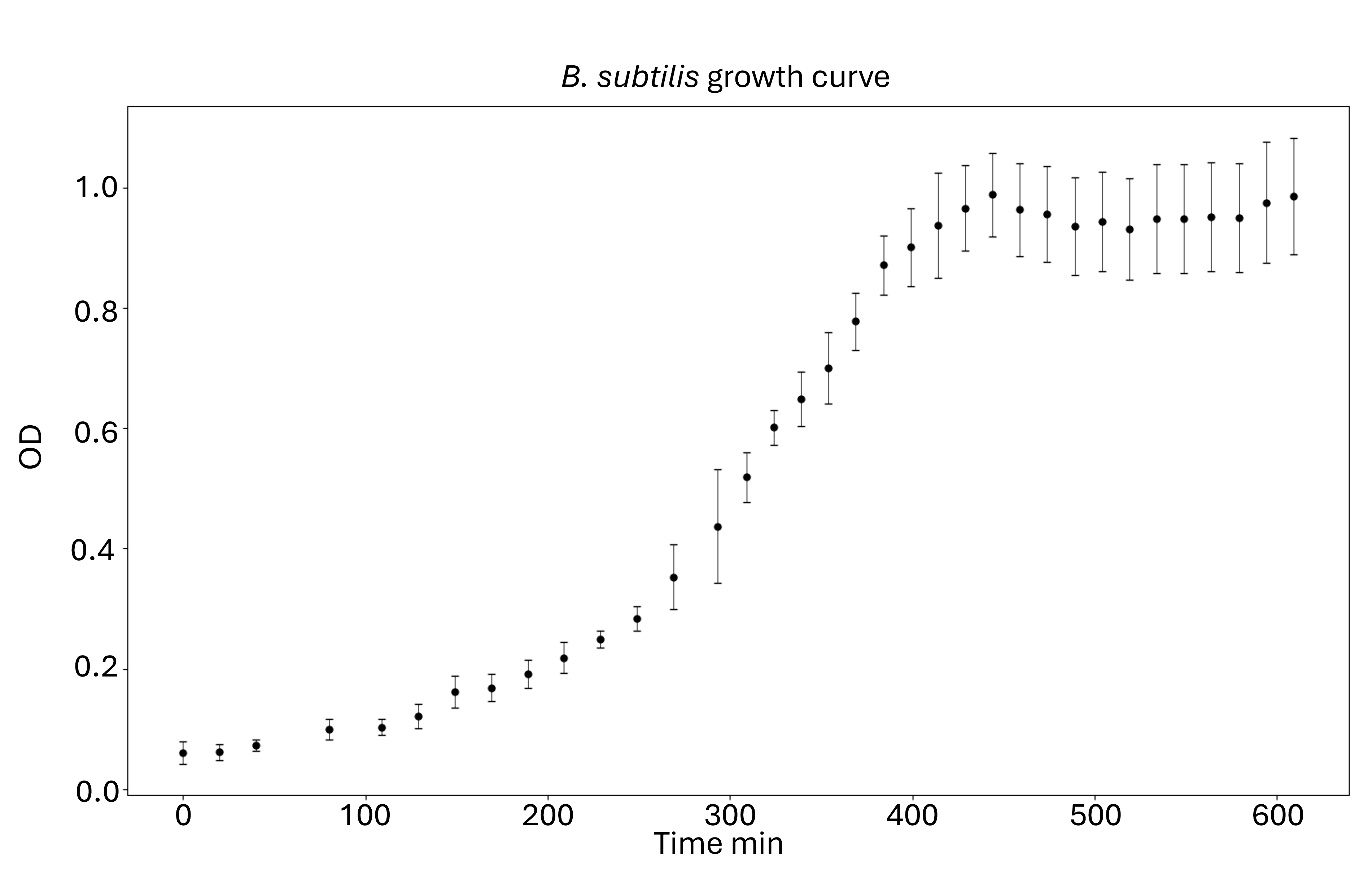}} \protect\caption{Growth curve of the B. subtilis strain used at $37^\circ\text{C}$.} 
\end{center}
\end{figure}

\begin{figure}[H]
\begin{center}
\resizebox{0.95\columnwidth}{!}
{\includegraphics{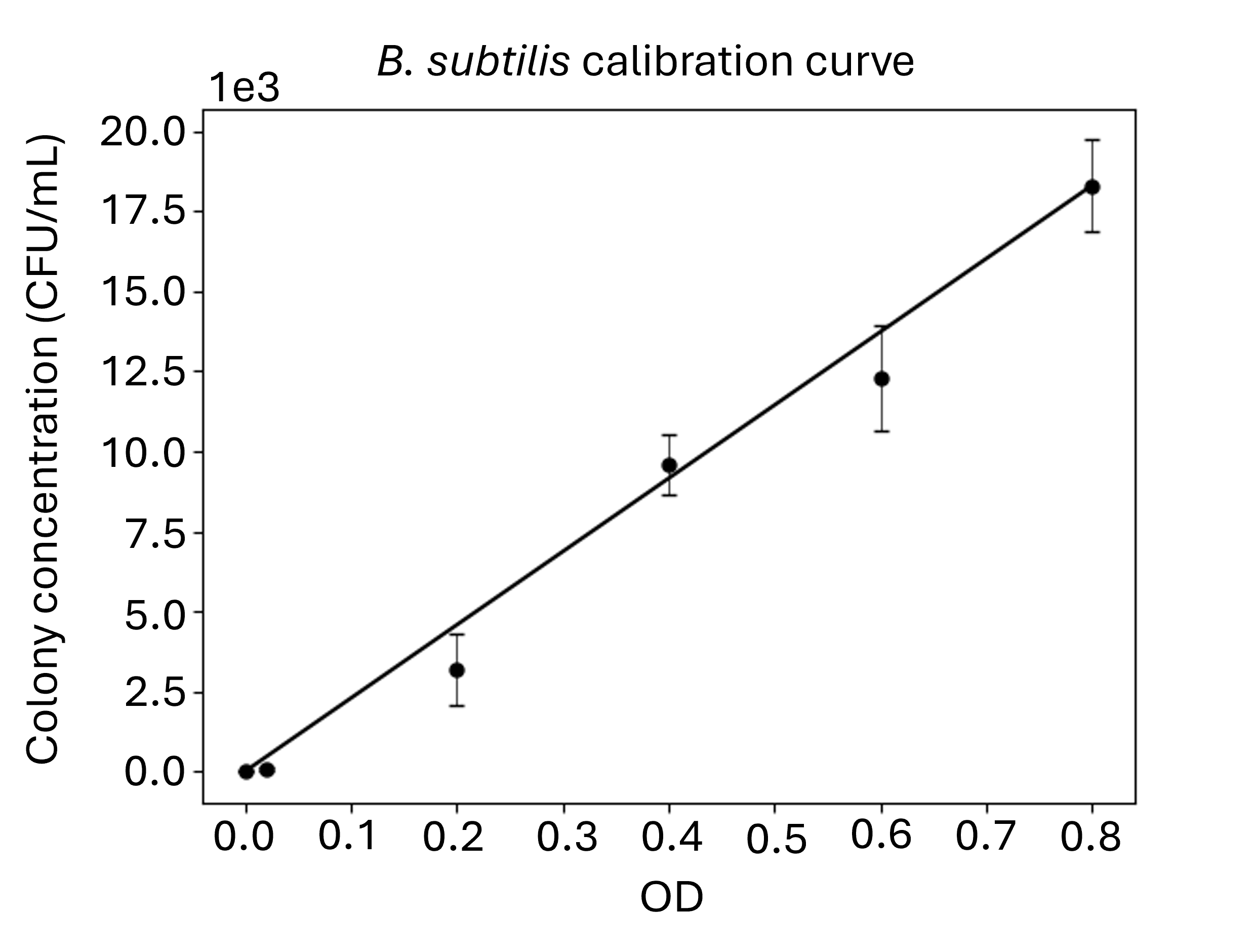}} \protect\caption{\textit{B. subtilis} calibration curve relating OD measurements and the corresponding number of CFUs in inoculated agar plates. } 
\end{center}
\end{figure}

\section{Experimental setup}

\begin{figure}[H]
\begin{center}
\resizebox{1\columnwidth}{!}
{\includegraphics{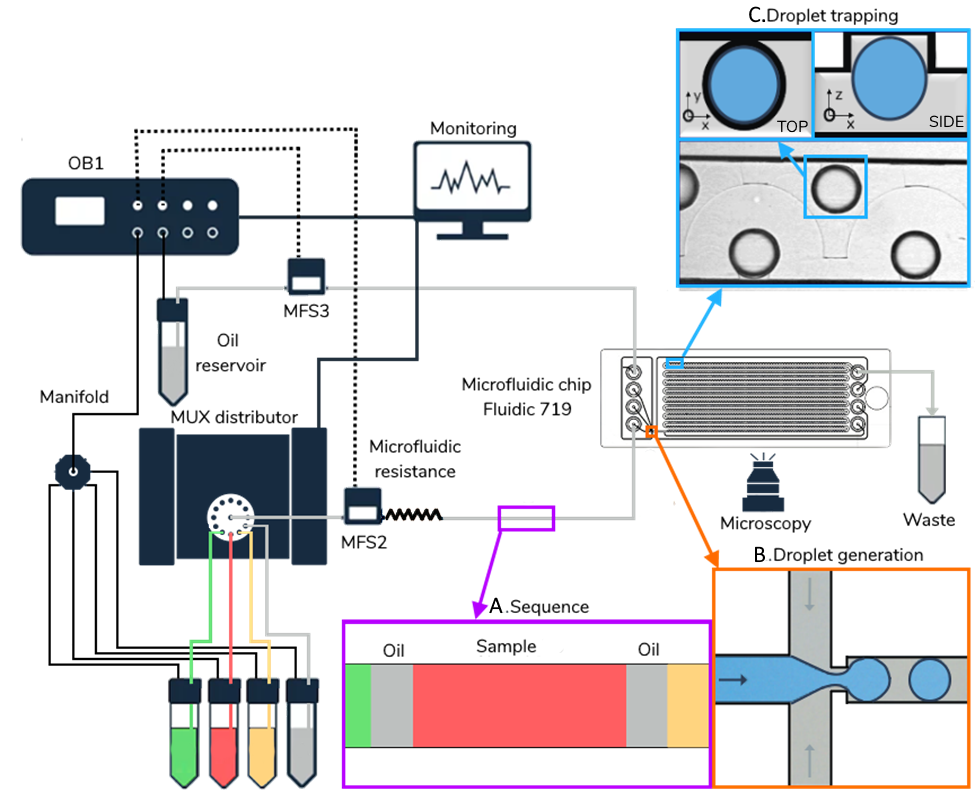}} \protect\caption{Representation of the microfluidic platform used in the experiments, illustrating the droplet production process in three key steps: A. Forming a dispersed phase sequence using the multi-injection valve. B. Producing droplets through flow control in the chip’s flow-focusing device. C. Stabilizing droplets in microfluidic traps along the channel.
} \label{fig:APN}
\end{center}
\end{figure}

\section{Droplet size distribution}

\begin{figure}[H]
\begin{center}
\resizebox{0.95\columnwidth}{!}
{\includegraphics{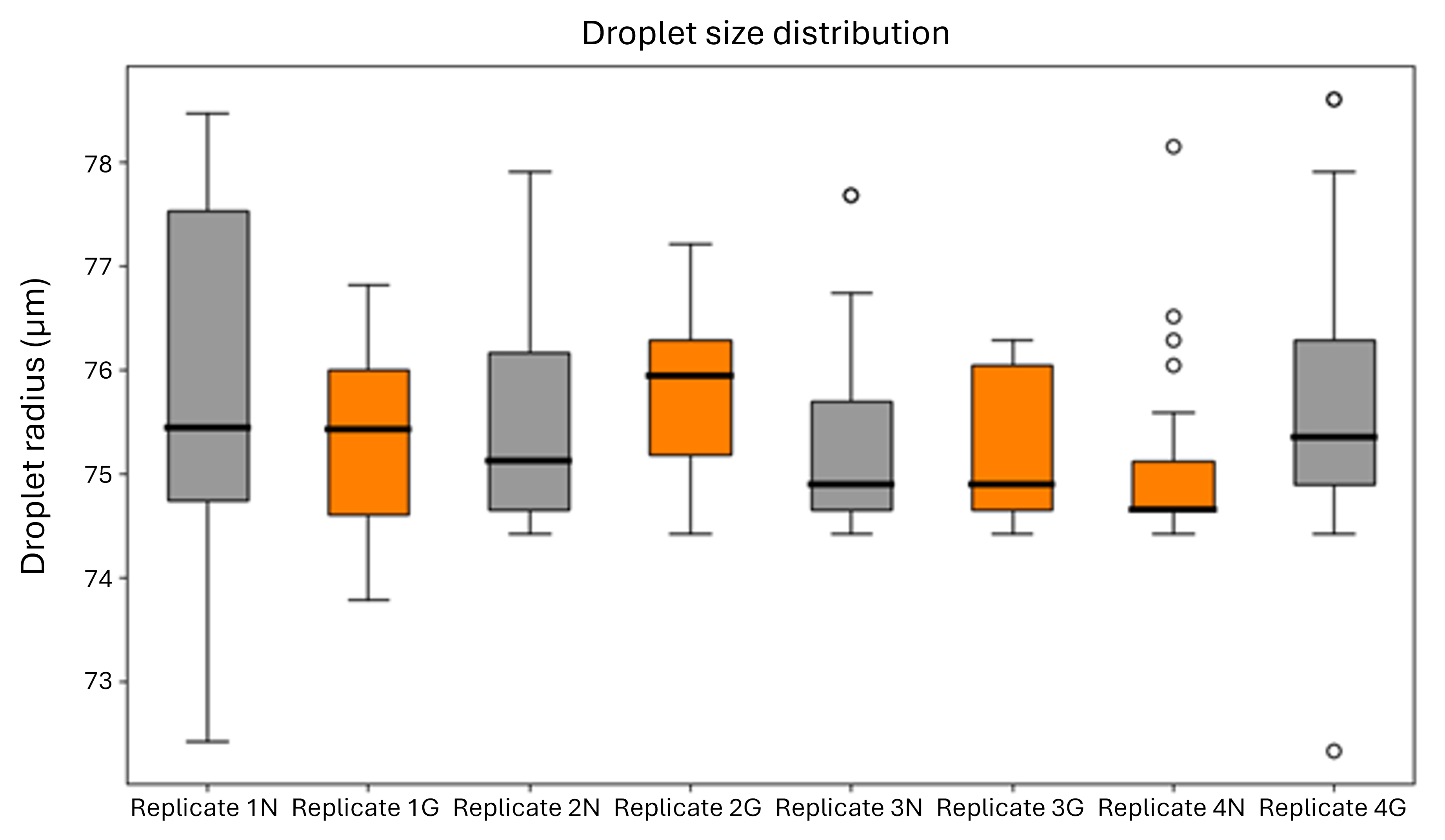}} \protect\caption{Droplet size distribution across different replicates and different droplet content. Four experiments were performed alternating production between droplets with and without glycerol. Droplets were monodisperse with a consistent radius suitable for trapping.} \label{fig:monodisperse}
\end{center}
\end{figure}

\section{Perfluorinated oil oxygen permeability evaluation.}

\begin{figure}[H]
\begin{center}
\resizebox{1\columnwidth}{!}
{\includegraphics{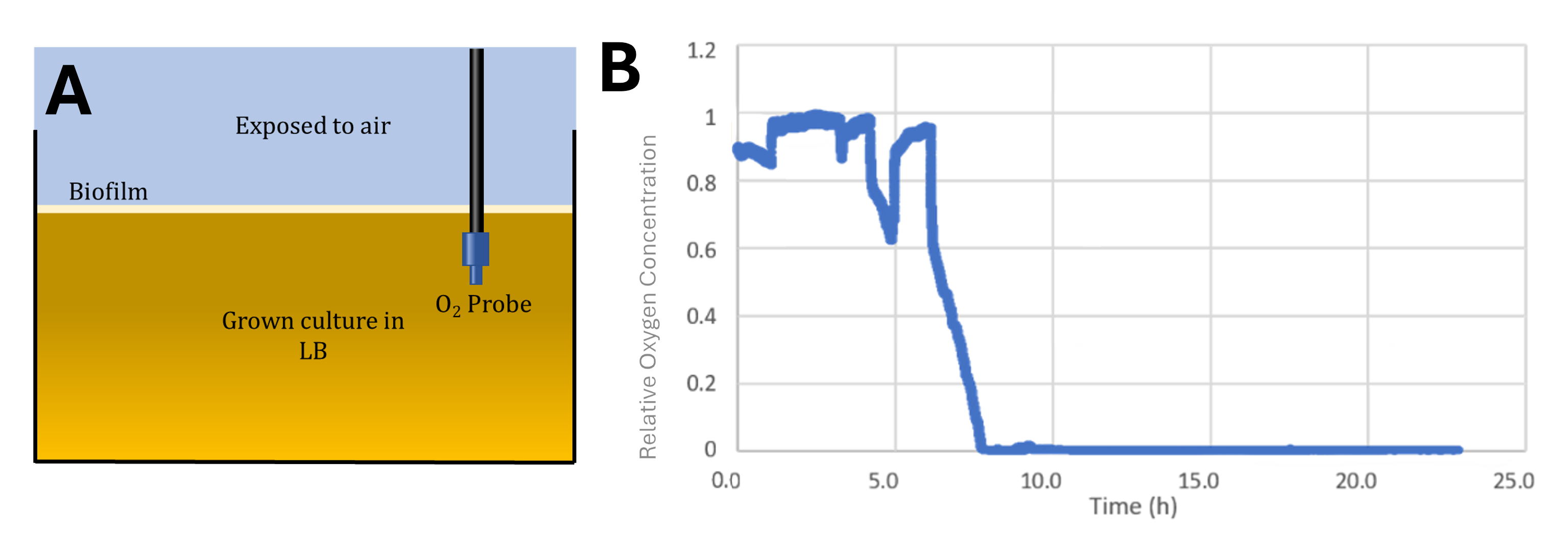}} \protect\caption{Oxygen concentration measurements in a standard liquid culture. A. The probe is submerged in LB medium, where bacteria grow and form a pellicle at the air-liquid interface. B. Oxygen levels remain near atmospheric values for the first six hours before dropping sharply and stabilizing.
} \label{fig:oxgy1}
\end{center}
\end{figure}

Oxygen concentration was monitored in various liquids under different conditions using a light sensor (OXY-1 SMA, PreSens Precision Sensing GmbH, Germany). The relative oxygen concentration, measured against atmospheric levels, was tracked in a growing \textit{B. subtilis} colony cultured in Lysogeny Broth (LB) medium exposed to air (Fig.~\ref{fig:oxgy1}A). A pellicle formed at the liquid surface, and after six hours, oxygen levels dropped sharply (Fig.~\ref{fig:oxgy1}B). This decrease can be attributed to the combined effects of the biofilm at the interface restricting oxygen diffusion into the liquid and the bacterial community's oxygen consumption outpacing the diffusion rate.

\begin{figure}[H]
\begin{center}
\resizebox{1\columnwidth}{!}
{\includegraphics{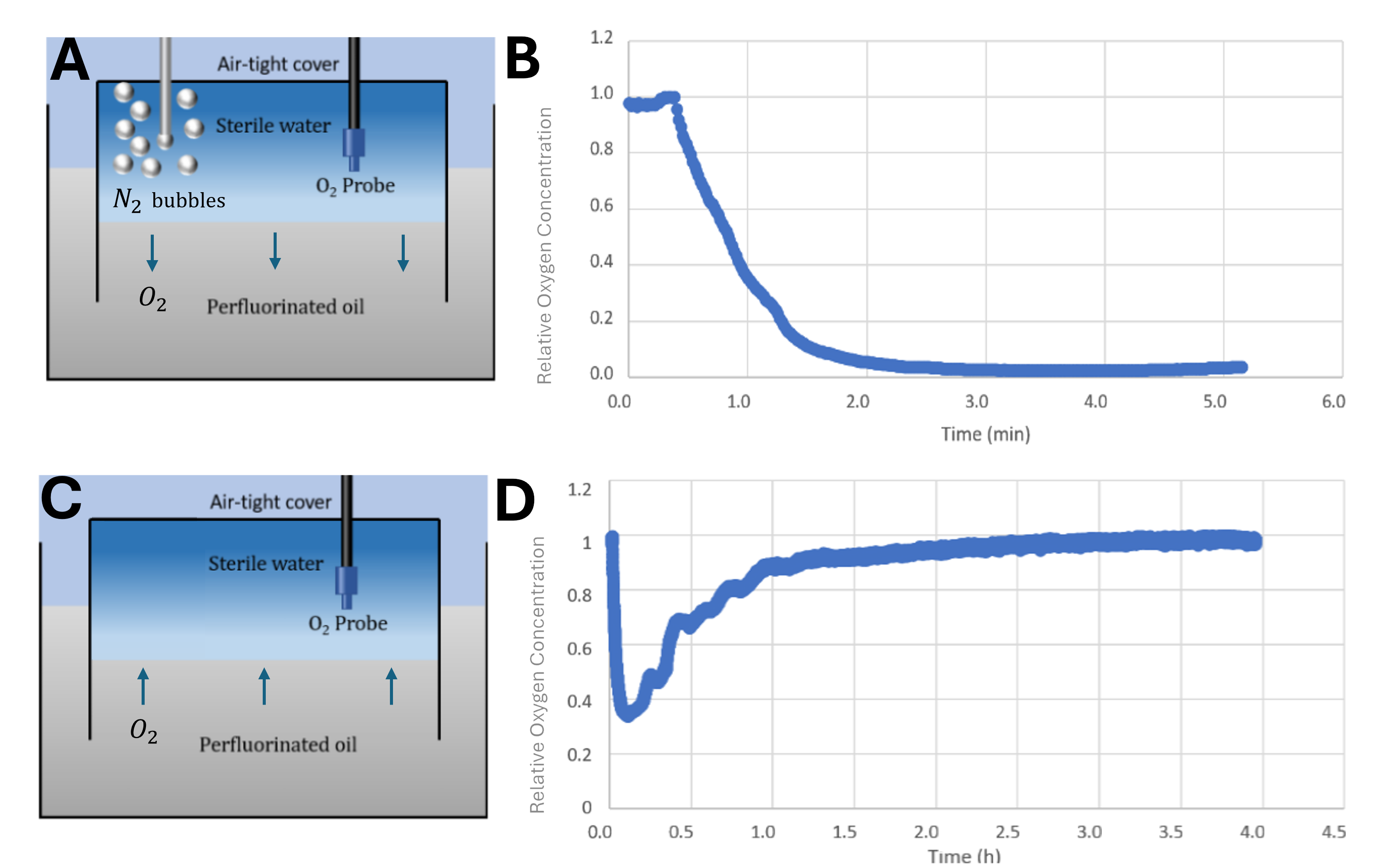}} \protect\caption{Oxygen Permeability Tests with Perfluorinated Oil: A. Nitrogen gas is bubbled into a water container, which is exposed to the environment only through an oil "window," to reduce oxygen concentration. B. During bubbling, oxygen is depleted within minutes. C. Once bubbling stops, oxygen begins to diffuse back into the water through the oil. D. The oxygen level gradually returns to the atmospheric value.
} \label{fig:oxy2}
\end{center}
\end{figure}

To verify oxygen diffusion from oil into water, an additional experiment was conducted. Oxygen concentration was measured in a sealed container with sterile water, connected to the exterior only through an oil "window." When nitrogen gas was bubbled through the system (Fig.~\ref{fig:oxy2}A), oxygen levels gradually declined until depletion (Fig.~\ref{fig:oxy2}B). After stopping the nitrogen flow (Fig.~\ref{fig:oxy2}C), oxygen levels rose back to atmospheric values, indicating diffusion through the oil (Fig.~\ref{fig:oxy2}D). These results demonstrate that the oil can serve as an oxygen reservoir, maintaining aerobic conditions for at least eight hours.

\begin{figure}[H]
\begin{center}
\resizebox{1\columnwidth}{!}
{\includegraphics{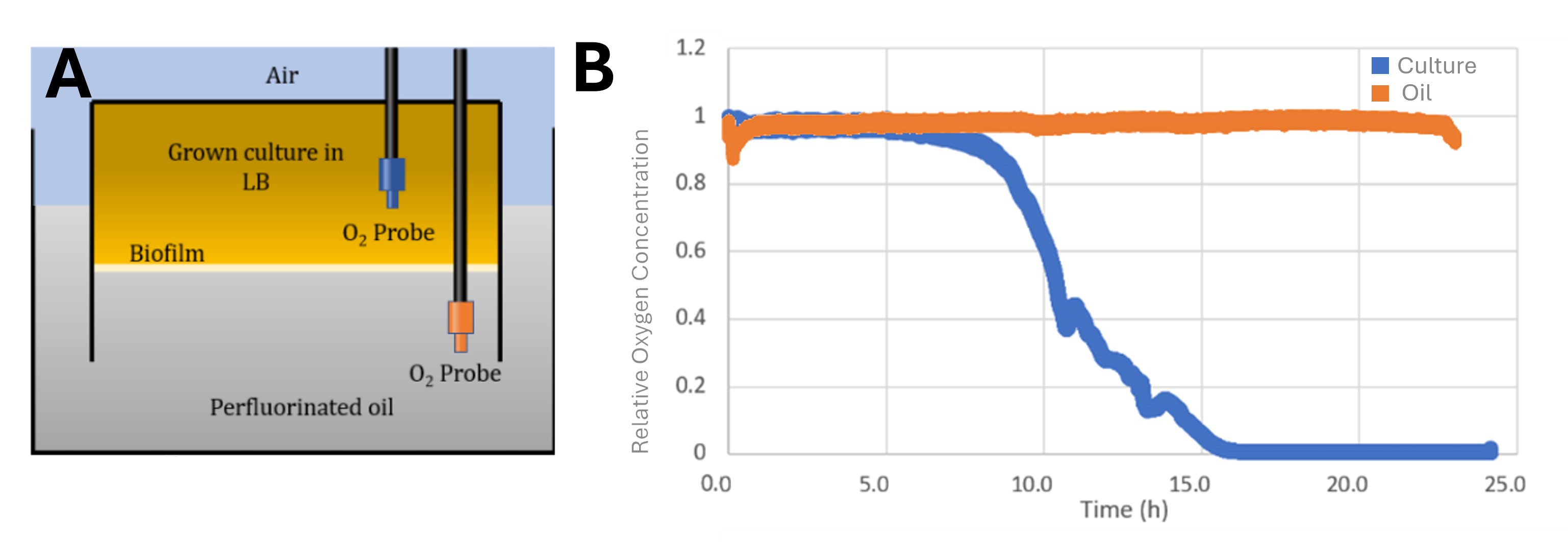}} \protect\caption{Finally, oxygen concentration was monitored in a bacterial culture, similar to the first experiment, but separated from the atmosphere by an oil layer (Fig.~\ref{fig:oxy3}A). Oxygen levels in the medium remained stable for the first six hours before declining (Fig.~\ref{fig:oxy3}B). In contrast, oxygen concentration within the oil remained constant throughout the experiment. This behavior was comparable to that observed in colonies exposed to air.
} \label{fig:oxy3}
\end{center}
\end{figure}

Finally, the oxygen concentration was monitored in a bacterial culture like the one in the first experiment and the oil separating it from the atmosphere (Fig.~\ref{fig:oxy3}A). The oxygen in the medium remained stable for 6 hours but dropped after 6 hours (Fig.~\ref{fig:oxy3}B). Meanwhile, the level of oxygen inside the oil was stable during the whole process. This behaviour was found comparable to that of the colonies exposed air.

These experiments lead to two main conclusions. First, the interface with perfluorinated oil induces pellicle formation in \textit{B. subtilis}, similar to the air-liquid interface, confirming its suitability for biofilm studies. Second, if oxygen levels decrease inside the droplets, oxygen from the oil will diffuse into partially compensate for the loss.

\section{Training of the Variational Autoencoder}

This section provides a detailed description of the training pipeline used for the Variational Autoencoder (VAE) models. We explain how input image crops were transformed into a rotation-invariant representation through a polar Fourier decomposition, and how this representation was used to train a VAE architecture tailored for unsupervised learning of local structural features in droplet images. 

\subsection*{Data Preprocessing and Angular Transformation}

From each brightfield droplet image (``Training droplets''), we extract 10{,}000 circular random crops (``Representative random crops'') of fixed diameter (24 px $\approx$ 1.86 $\mu$m). A representation of these images and their crops is shown in Fig.~8A.

To make the model insensitive to the orientation of bacterial structures, each crop is transformed into a representation that captures radial and angular patterns regardless of how the crop is rotated. This is achieved by converting the image into polar coordinates and analyzing its frequency content along radial and angular directions. The transformation consists of the following conceptual steps:

\begin{enumerate}
    \item \textbf{Polar Coordinates and Flattening.}
    \begin{itemize}
        \item For a given crop, we compute three flattened arrays of length $N$: $\{ f(p) \}_{p=1}^N$, $\{ r(p) \}_{p=1}^N$, and $\{ \omega(p) \}_{p=1}^N$, where $f(p)$ is the grayscale intensity of pixel $p$, $r(p)$ is its radial distance from the center of the crop, and $\omega(p)$ is its polar angle (measured from the positive x-axis).
    \end{itemize}

    \item \textbf{Defining the Concentric Radial Bins.}
    \begin{itemize}
        \item Let $R_{\text{max}} = \max_p r(p)$ be the outer radius of the crop.
        \item Define the boundaries of the six radial bins as $r_0, r_1, \ldots, r_6$, with $r_0 = 0$ and $r_6 = R_{\text{max}}$, where $r_i = \frac{i}{6} R_{\text{max}}$ for $i = 0, \ldots, 6$. Note that $r_i$ are real-valued and not constrained to integers, since radial distances are based on pixel distances from the center, which can be fractional.
        \item Assign each pixel $p$ to a radial-bin index $j \in \{1, \ldots, 6\}$ according to the interval $r(p) \in [r_{j-1}, r_j)$, so pixel $p$ belongs to bin $j$.
    \end{itemize}

    \item \textbf{Angular Frequencies: Integration over Each Radial Ring.}
    \begin{itemize}
        \item We fixed $n = 8$ angular frequencies, indexed by $k = 0, 1, \ldots, 7$, which provided a good balance between capturing meaningful angular structure and keeping the representation compact and stable. For each radial bin $j$ and each frequency $k$, we compute:
        \[
        u_{k,j} = \int_{\{ p : p \text{ in bin } j \}} f(p) \cdot e^{i k \omega(p)} \, d\omega
        \]
        \item To approximate this integral, sort the set $\{(\omega(p), f(p)) : \text{inds}[p] = j\}$ by increasing $\omega$ and apply the trapezoidal rule:
        \[
        u_{k,j} \approx \sum_{\ell=1}^{N_j-1} \frac{f(\ell) e^{i k \omega(\ell)} + f(\ell+1) e^{i k \omega(\ell+1)}}{2} \cdot \Delta \omega_\ell
        \]
        where $N_j$ is the number of pixels in bin $j$ and $\Delta \omega_\ell = \omega(\ell+1) - \omega(\ell)$.
        \item For $k = 0$, this simplifies to:
        \[
        u_{0,j} \approx \sum_{p : r(p) \in \text{bin } j} f(p) \, \Delta \omega(p)
        \]
        which represents the total intensity (DC component) in radial bin $j$.
        \item For $k > 0$, $u_{k,j}$ captures the strength of the $k$-th angular variation within each ring.
    \end{itemize}

    This process captures how brightness varies with angle at each radius, allowing the model to detect rotational features such as circular textures, rings, or symmetric patterns.

    \item \textbf{Rotation Invariance via Magnitudes.}
    \begin{itemize}
        \item If the crop is rotated by an angle $\Delta \omega_0$, each coefficient transforms as $u_{k,j} \rightarrow u_{k,j} e^{i k \Delta \omega_0}$.
        \item By discarding the complex phase and retaining only $|u_{k,j}|$, the magnitude matrix $|U| \in \mathbb{R}^{6 \times 8}_{\geq 0}$ becomes invariant to any in-plane rotation.
    \end{itemize}

    This step ensures that crops with the same structure but different orientations are represented identically, so the model focuses on structure rather than absolute orientation. 
    The resulting 48-dimensional input vector $|u_{k,j}|$ (for $k = 0, \ldots, 5$ and $j = 1, \ldots, 8$) is reshaped into 6 channels (radial bins) $\times$ 8 positions (angular frequencies) and processed by the VAE encoder, a representation of this vector is shown in Fig.~8A.
\end{enumerate}

\subsection*{VAE Architecture and Latent-Space Visualization}

To extract meaningful features from these representations, we used a Variational Autoencoder (VAE) — a neural network model designed to learn compressed, low-dimensional encodings of input data. The VAE consists of an encoder, which maps the input into a latent space, and a decoder, which attempts to reconstruct the input from this latent representation. Although the decoder is included during training to support learning, only the encoder and its latent output are used in our analysis.

The VAE architecture is composed of the following layers:
\begin{itemize}
    \item Conv1D (6 $\rightarrow$ 32): Convolution along the angular axis
    \item Conv1D (32 $\rightarrow$ 64): Feature channel expansion
    \item Conv1D (64 $\rightarrow$ 64): Channel refinement
    \item Flatten $\rightarrow$ 384-dimensional vector
    \item Two parallel fully connected layers output the mean $\mu$ and log-variance $\log \sigma^2$ of a univariate Gaussian. 
\end{itemize}
A representation of the VAE architecture is shown in Fig.~8B.

A single latent variable $z \sim \mathcal{N}(\mu, \sigma^2)$ is sampled. The VAE is trained by minimizing the sum of the reconstruction error and the Kullback-Leibler divergence between the approximate posterior $q(z \mid \mu, \sigma^2)$ and the prior $\mathcal{N}(0, 1)$.

To visualize the learned latent space, we collect $z$ values from the same set of random crops used during training, plot a histogram, and select representative bins (e.g., 1, 4, 7, 10, 13, 16, 19, 22, 25, 28). Below the histogram, circular insets show example crops from each bin. As $z$ increases, the crops show a progression from uniform intensity to more complex textures. A representation of the histogram and the circular crops is shown in Fig.~8B.

\medskip
\noindent

\begin{figure}[H]
\begin{center}
\resizebox{1\columnwidth}{!}
{\includegraphics{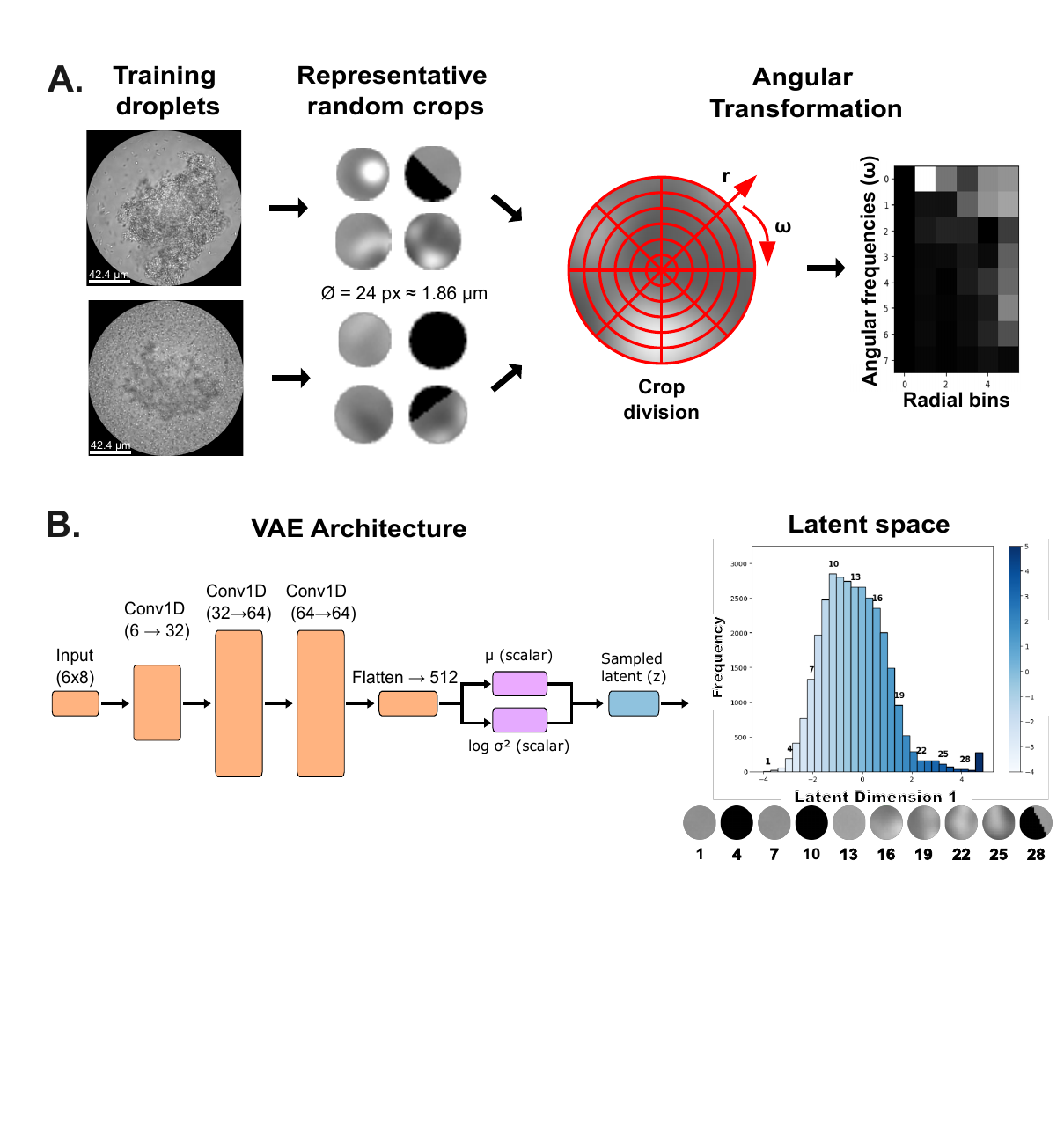}} \protect\caption{A. Angular‐frequency processing. Top panels: two representative brightfield droplet images used for training. From each image, we extract multiple circular random crops (diameter 24 px \(\approx\) 1.86 µm). This size was chosen to capture relevant features such as individual cells. A few example crops are shown. Each crop is divided into six concentric radial bins and eight equal‐angle wedges generating a \(6\times 8\) angular‐frequency matrix.
B. VAE architecture and latent‐space visualization. The \(6\times 8\) matrix of real‐valued angular frequencies is passed through a VAE encoder consisting of three 1D convolutional layers (Conv1D: \(6\to32\), \(32\to64\), \(64\to64\)), followed by flattening into a 512‐dimensional vector and two parallel dense layers that output the scalar mean \(\mu\) and log‐variance \(\log\sigma^2\). A single latent variable \(z\sim \mathcal{N}(\mu,\sigma^2)\) is sampled. The histogram on the right shows the distribution of \(z\) values (latent dimension 1) for sampled crops at labeled bins 1, 4, 7, 10, 13, 16, 19, 22, 25, 28 illustrating a gradual transition from low‐contrast uniform patches (low \(z\)) to high‐contrast patterns (high \(z\)).%
}
 \label{fig:SI_8}
\end{center}
\end{figure}

\thispagestyle{empty}